\documentclass[12pt]{article}

\usepackage[utf8]{inputenc}
\usepackage[T1]{fontenc}
\usepackage{lmodern}
\usepackage{amsmath,amssymb}
\usepackage{amsthm}
\usepackage{dsfont}
\usepackage{mathrsfs}
\usepackage{float}
\usepackage{graphicx}
\usepackage{babel}
\usepackage{natbib}
\usepackage{setspace}
\theoremstyle{plain}
\newtheorem{assumption}{\protect\assumptionname}
\theoremstyle{plain}
\newtheorem{prop}{\protect\propositionname}
\theoremstyle{definition}
 
\theoremstyle{remark}
\newtheorem{rem}{\protect\remarkname}
\theoremstyle{plain}
\newtheorem{cor}{\protect\corollaryname}
\theoremstyle{plain}
\newtheorem{thm}{\protect\theoremname}

\providecommand{\assumptionname}{Assumption}
\providecommand{\corollaryname}{Corollary}
\providecommand{\examplename}{Example}
\providecommand{\propositionname}{Proposition}
\providecommand{\remarkname}{Remark}
\providecommand{\theoremname}{Theorem}

\usepackage[top=1in, bottom=1in, left=1in, right=1in]{geometry}

\title{Debiased Machine Learning when Nuisance Parameters Appear in Indicator Functions\thanks{I would like to thank Toru Kitagawa, Susanne Schennach, Andriy Norets, Soonwoo Kwon, Jonathan Roth, and Peter Hull for helpful comments. I gratefully acknowledge financial support from Department of Economics, Brown University (Merit Dissertation Fellowship).}}
\author{Gyungbae Park\thanks{School of Management and Economics, Chinese University of Hong Kong, Shenzhen}}
\date{October 8th, 2026}

\begin{document}
\maketitle

\begin{abstract}
This paper studies debiased machine learning when nuisance parameters appear in indicator functions. An important example is maximized average welfare gain under optimal treatment assignment rules. For asymptotically valid inference for a parameter of interest, the current literature on debiased machine learning relies on Gateaux differentiability of the functions inside moment conditions, which does not hold when nuisance parameters appear in indicator functions. In this paper, we propose smoothing the indicator functions, and develop a bias-aware asymptotic distribution theory for this class of models. The asymptotic behavior of the proposed estimator exhibits a trade-off between nuisance estimation error and the negative approximation bias due to smoothing. We study how a parameter which controls the degree of smoothing can be chosen optimally to minimize an upper bound of the asymptotic mean squared error. A Monte Carlo simulation supports the asymptotic distribution theory, and an empirical example illustrates the implementation of the method.
\end{abstract}

\vspace{1em}

\section{Introduction}
\label{sec_intro}

This paper studies debiased machine learning (DML) when nuisance parameters appear in indicator functions. An important example is the maximized average welfare gain under optimal treatment assignment rules. Provided that unconfoundedness assumption holds, the conditional average treatment effect (CATE) function is identified. The parameter of interest is represented by the expectation of a moment function where the moment function consists of indicator functions. For asymptotically valid inference for a parameter of interest, the current literature on debiased machine learning relies on Gateaux differentiability of the functions inside moment conditions. However, Gateaux differentiability does not hold when nuisance parameters appear in indicator functions, which makes the development of valid inference procedures an open problem.

Let $W=\left(Y,X^{'}\right)^{'}$ denote an observation where $Y$ is an outcome variable with a finite second moment and $X$ is a high-dimensional vector of covariates. Let
\[
\boldsymbol{\gamma}_{0}\left(x\right)\equiv\mathbb{E}\left[Y\mid X=x\right]
\]
be the conditional expectation of $Y$ given $X\in\mathcal{X}.$ Let $\boldsymbol{\gamma}:\mathcal{X}\rightarrow\mathbb{R}$ be a function of $X.$ Define $m\left(w,\boldsymbol{\gamma}\right)$ as a function of the function $\boldsymbol{\gamma}$ (i.e. a functional of $\boldsymbol{\gamma}$), which depends on an observation $w.$ The parameter of interest $\theta_{0}$ has the following expression:
\[
\theta_{0}=\mathbb{E}\left[m\left(W,\boldsymbol{\gamma}_{0}\right)\right].
\]
\citet{Chernozhukov.et.al.2022a} shows an asymptotic distribution theory for DML when $m\left(w,\boldsymbol{\gamma}\right)$ is linear or nonlinear in $\boldsymbol{\gamma}.$ When $m\left(W,\boldsymbol{\gamma}\right)$ is nonlinear in $\boldsymbol{\gamma},$ \citet{Chernozhukov.et.al.2022a} linearizes it and extends results for the linear case to the linearized function. The key assumption is that the remainder in the linearization, employing Gateaux differentiability in a neighborhood of the true parameter, is bounded by a constant. This assumption is crucial in showing asymptotic normality of the DML estimator in the nonlinear case as it renders the remainder term negligible. On the other hand, we focus on problems where $\boldsymbol{\gamma}$ appears in indicator functions, and hence $m\left(w,\boldsymbol{\gamma}\right)$ is not Gateaux differentiable in $\boldsymbol{\gamma}.$ This motivates us to propose an alternative approach in which a smoothing function is used to smooth the indicator function.

DML has been widely studied in econometrics literature. \citet{Chernozhukov.et.al.2017} and \citet{Chernozhukov.et.al.2018} propose a general DML approach for valid inference in the context of estimating causal and structural effects. \citet{Semenova.and.Chernozhukov2020} studies DML estimation of the best linear predictor (approximation) for structural functions including conditional average structural and treatment effects. Recently, \citet{Chernozhukov.et.al.2022a} proposes automatic DML for linear and nonlinear functions of regression equations. They provide the average policy effect, weighted average derivative, average treatment effect and the average equivalent variation bound as examples of linear functions of regression equations. As an example of a nonlinear function, they discuss the causal mediation analysis of \citet{Imai.et.al.2010}. \citet{Chernozhukov.et.al.2022b} derives the convergence rate and asymptotic results for linear functionals of regression equations. \citet{Chernozhukov.et.al.2022c} proposes a general method to construct locally robust moment functions for generalized method of moments estimation. The two key features of DML are orthogonal moments functions and cross-fitting. (Neyman) Orthogonal moment functions are used to mitigate regularization and/or model selection bias, and to avoid using plug-in estimators. When employing regularized machine learners in causal or structural estimation, squared bias may decrease at a slower rate than variance. As a result, confidence interval coverage can be poor and estimators will not be $\sqrt{n}$-consistent (\citet{Chernozhukov.et.al.2017}, \citet{Chernozhukov.et.al.2018}, and \citet{Chernozhukov.et.al.2022c}). Combining orthogonal moment functions with cross-fitting makes inference available when regression estimators are high-dimensional.

This paper contributes to the expanding DML literature by considering estimation and inference for non-differentiable functions. We propose smoothing the indicator functions, and develop an asymptotic distribution theory for a class of models which involves indicator functions. We introduce a sigmoid function to smooth the indicator function, and a smoothing parameter which controls the smoothness of the sigmoid function. The asymptotic expansion contains a Gaussian sampling component, a potentially non-negligible second-order nuisance-estimation error, and a negative one-sided smoothing approximation bias. The approximation bias is negative because the smoothed target lies weakly below the original target. The smoothing parameter governs a trade-off between the latter two components: making the sigmoid approximation steeper reduces the approximation error but increases its sensitivity to nuisance-estimation error. Our analysis imposes no margin condition and therefore accommodates distributions that place probability mass at or arbitrarily close to the cutoff. We propose an oracle smoothing parameter that balances worst-case bounds on the two errors and minimizes a conservative upper bound on the leading asymptotic mean squared error. Because these errors need not vanish relative to sampling uncertainty, conventional inference may be invalid. Building on the bias-aware approach of \citet{Armstrong.and.Kolesar2020}, we construct an oracle confidence interval based on a folded-normal critical value that accounts for the maximum absolute drift and establish its pointwise asymptotic validity.

We conduct Monte Carlo simulations to verify our theoretical results, and an empirical analysis to illustrate implementation of the methods. The simulation results illustrate the finite-sample behavior of the proposed estimator and the bias-aware confidence interval when the smoothing parameter is chosen according to the proposed rule. In our empirical analysis, we use experimental data from the National Job Training Partnership Act (JTPA) Study (\citet{Bloom.et.al.1997}). When the smoothing parameter is chosen optimally, we find that the estimate of maximized welfare gain is similar to previous studies.

Non-differentiability often arises in causal inference problems involving treatment assignment rules. \citet{D'Adamo.2022} studies the estimation of optimal treatment rules under partial identification. \citet{Christensen.et.al.2023} estimates treatment rules under directional differentiability with respect to a finite-dimensional nuisance parameter. Many papers propose various approaches to handle non-differentiability in specific settings. \citet{Horowitz1992} uses the sigmoid function to smooth the indicator function in analyzing the binary response model. The parameters of interest in \citet{Horowitz1992} are coefficients of the single index model, while our parameter of interest is the value of a criterion function. \citet{Zhou.et.al2017} replaces the 0-1 loss with the smoothed ramp loss in the framework of residual weighted learning to estimate individualized treatment rules. They construct the smoothed ramp loss by replacing the sharp cutoff on the interval $\left[-1,1\right]$ with a quadratic smoothing function. In our approach, we employ a smoothing parameter that depends on the sample size to smooth the indicator function using the sigmoid function. On the other hand, \citet{Chen.et.al.2003} studies a class of semiparametric optimization estimators when criterion functions are not smooth, and does not introduce a smoothing function when deriving the asymptotic distribution. \citet{Hirano.and.Porter.2012} shows that if the target object is non-differentiable in the parameters of the data distribution, there exist no estimator sequences that are locally asymptotically unbiased. Non-differentiability also arises in the nonparametric IV quantile regression through the non-smooth generalized residual functions. In response, \citet{Chen.and.Pouzo.2012} proposes a class of penalized sieve minimum distance estimators. \citet{Levis.et.al.2023} studies the covariate-assisted version of the Balke and Pearl bounds (\citet{Balke.and.Pearl.1997}), which are characterized as non-smooth functionals (specifically, a max function). They smooth the max function using the log-sum-exp (LSE) function and provide an estimator based on the nonparametric efficient influence function for the smoothed functional. In contrast, we smooth the indicator function using a sigmoid function.

Standard bootstrap consistency fails in the presence of non-differentiability, which has led researchers to propose alternative bootstrap methods. \citet{Andrews2000} shows inconsistency of the bootstrap if the parameter lies on the boundary of a parameter space defined by linear or nonlinear inequality constraints. \citet{Fang.and.Santos.2018} studies inference for (Hadamard) directionally differentiable functions, and \citet{Hong.and.Li.2018} proposes a numerical derivative based Delta method to show consistent inference for functions of parameters that are only directionally differentiable. Recently, \citet{Kitagawa.et.al2020} characterizes the asymptotic behavior of the posterior distribution of functions which are locally Lipschitz continuous but possibly non-differentiable. 

Various works study inference for welfare under optimal treatment assignment rules. \citet{Whitehouse.et.al.2026} use softmax smoothing and debiasing to conduct inference on optimal policy values and other maximum-type functionals. For binary treatments, their softmax approximation is equivalent, up to the baseline outcome, to the sigmoid approximation used here. Our DML estimator employs an orthogonal moment function that incorporates a Riesz representer for the expectation of the derivative of the smoothing function. As \citet{Chernozhukov.et.al.2022a} point out, an orthogonal moment function containing a Riesz representer can be interpreted as the efficient influence function for the smoothed functional. Whitehouse et al. allow arbitrarily large probability mass at exact treatment ties, but impose a polynomial density, or soft margin, condition on nonzero treatment gaps near the cutoff. This condition makes the smoothing bias asymptotically negligible and permits centered normal inference at the parametric rate. \citet{Chen.et.al.2025} take a different approach and obtain inference for optimal welfare and more general value functionals using nonparametric plug-in estimators without smoothing. For optimal welfare, their parametric-rate result relies on a regular level set condition that, together with their continuous-covariate setting, implies that the CATE equals zero with probability zero. \citet{Luedtke.and.van.der.Laan2016} study inference for the mean outcome under optimal treatment rules by developing a regular and asymptotically linear estimator. \citet{Semenova2023a} and \citet{Semenova2024} study estimation and inference for objects involving maxima or minima of nuisance functions. These works consider plugging machine learning estimates of the nuisance functions directly into non-differentiable functionals under margin restrictions. In contrast, our analysis imposes neither a regular level set condition nor a margin restriction on nonzero CATE values near the cutoff. It therefore accommodates both positive probability mass at zero and unrestricted concentration of nonzero CATE values arbitrarily close to zero. Positive mass at the cutoff is explicitly reflected in the limiting variance. Because the smoothing and nuisance-estimation errors may remain asymptotically non-negligible, we select an oracle smoothing parameter by minimizing a conservative upper bound on the leading asymptotic mean squared error and use an upper bound on their combined drift to construct an oracle Armstrong-Kolesár confidence interval with pointwise asymptotic coverage. This provides policymakers with an inference strategy under weaker restrictions on the distribution of the CATE near the cutoff. \citet{Kitagawa.and.Tetenov2018} proposes a resampling based inference procedure for optimized welfare with potentially conservative coverage. \citet{Andrews.et.al2023} studies estimators and confidence intervals for the welfare at an estimated policy that controls the winner’s curse.

In addition, the average welfare gain from the unrestricted optimal treatment plays a critical benchmarking role in policy learning. \citet{Manski2004} evaluates the performance of statistical treatment rules in terms of their maximum regret and provides finite‐sample regret bounds for conditional empirical success (CES) rules. \citet{Kitagawa.and.Tetenov2018} assesses the properties of estimated treatment rules by their average welfare regret relative to the maximum feasible welfare gain by using the empirical welfare maximization method. \citet{Athey.and.Wager2021} studies policy learning using observational data with the doubly-robust approach from \citet{Chernozhukov.et.al.2022c}. While much of the literature focuses on identifying the optimal treatment rule within a restricted class, the unrestricted optimal policy gain quantifies the maximum achievable benefit if treatments were allocated perfectly according to true conditional average treatment effects. \citet{Chernozhukov.et.al.2024a} views it as a measure of the heterogeneity of treatment effects which quantifies the potential improvement over the average effect achievable through optimally tailored treatment assignments. Our estimator provides a consistent measure of the average welfare gain under the unrestricted optimal treatment rule, suggesting that our estimated benchmark can then be used to compare with the average welfare achieved by restricted treatment rules. The difference between the two may serve as an indication of the welfare loss incurred when policies are restricted to a particular class, thereby quantifying the potential benefit of allowing for more tailored treatment assignments.

The rest of the paper is organized as follows. Section \ref{sec_nondiff} introduces maximized average welfare gain under optimal treatment rules as an example where the parameter of interest is a non-differentiable function of regression equations. Section \ref{sec_est} presents the estimation method and inference. Section \ref{sec_sim} gives simulation results. Section \ref{sec_emp} presents an empirical example. Section \ref{sec_conc} concludes the paper.

\section{Non-differentiable Effects}
\label{sec_nondiff}

In some cases the parameter of interest is the expectation of a non-differentiable function of regression equations. An important example is maximized average welfare gain under optimal treatment assignment rules. Consider the following potential outcomes framework. Let $D$ be a binary treatment status indicator and $Y\left(D\right)$ be a potential outcome. $Y\left(1\right)$ denotes the potential outcome upon receipt of the treatment, and $Y\left(0\right)$ represents the potential outcome without receipt of the treatment. The observed outcome $Y$ is written as
\[
Y=Y\left(D\right)=DY\left(1\right)+\left(1-D\right)Y\left(0\right)
\]
Let $W=\left(Y,X^{'}\right)^{'}$ denotes an observation, with $X=\left(D,Z^{'}\right)^{'}$ where $Z$ is a high-dimensional vector of covariates. High-dimensional covariates are often considered in recent causal inference literature including \citet{Semenova.and.Chernozhukov2020}. Let $\delta\left(Z\right)\in\left\{ 0,1\right\} $ be a treatment assignment function, where $\delta\left(Z\right)=1$ if treatment is assigned and $\delta\left(Z\right)=0$ if not. The maximized average welfare with respect to $\delta\left(Z\right)$ is expressed as follows.
\[
\mathbb{E}\left[Y\left(1\right)\delta\left(Z\right)+Y\left(0\right)\left(1-\delta\left(Z\right)\right)\right].
\]
Under the unconfoundedness assumption $\left(Y\left(1\right),Y\left(0\right)\right)\perp D\mid Z$ and the overlap condition, we can identify the CATE function
\[
\tau\left(Z\right)=\mathbb{E}\left[Y\left(1\right)-Y\left(0\right)\mid Z\right].
\]
If $\tau\left(Z\right)$ is known, the optimal treatment assignment rule $\delta^{*}\left(Z\right)$ is
\[
\delta^{*}\left(Z\right)=\mathds{1}\left\{ \tau\left(Z\right)>0\right\} .
\]
The welfare gain relative to the no-one treated policy is
\begin{eqnarray*}
\mathbb{E}\left[Y\left(1\right)\delta^{*}\left(Z\right)+Y\left(0\right)\left(1-\delta^{*}\left(Z\right)\right)\right]-\mathbb{E}\left[Y\left(0\right)\right] & = & \mathbb{E}\left[\left\{ Y\left(1\right)-Y\left(0\right)\right\} \mathds{1}\left\{ \tau\left(Z\right)>0\right\} \right]\\
 & = & \mathbb{E}\left[\tau\left(Z\right)\mathds{1}\left\{ \tau\left(Z\right)>0\right\} \right]
\end{eqnarray*}
where the second equality holds by the law of the iterated expectations. The parameter of interest can be expressed as $\theta_{0}=\mathbb{E}\left[m\left(W,\boldsymbol{\gamma}_{0}\right)\right]$ with
\begin{eqnarray*}
m\left(W,\boldsymbol{\gamma}\right) & = & \left[\gamma_{1}\left(X\right)-\gamma_{2}\left(X\right)\right]\mathds{1}\left\{ \gamma_{1}\left(X\right)-\gamma_{2}\left(X\right)>0\right\} \\
\boldsymbol{\gamma} & = & \left(\gamma_{1}\left(X\right),\gamma_{2}\left(X\right)\right)^{'}\\
\gamma_{1}\left(X\right) & = & \mathbb{E}\left[Y\mid Z,D=1\right]\\
\gamma_{2}\left(X\right) & = & \mathbb{E}\left[Y\mid Z,D=0\right].
\end{eqnarray*}
\citet{Hirano.and.Porter.2012} shows impossibility results for the estimation of non-differentiable functionals of the data distribution. In particular, when the target object is non-differentiable in the parameters of the data distribution, there exist no sequence of estimators that achieves local asymptotic unbiasedness. Even though $m\left(W,\boldsymbol{\gamma}\right)$ is non-differentiable in $\boldsymbol{\gamma},$ $\mathbb{E}\left[m\left(W,\boldsymbol{\gamma}\right)\right]$ is not necessarily non-differentiable. For example, if $\tau\left(Z\right)$ follows a normal distribution with mean $\mu$ and variance $\sigma^{2},$ we have
\begin{eqnarray*}
\theta_{0} & = & \mathbb{E}\left[m\left(W,\boldsymbol{\gamma}_{0}\right)\right]\\
 & = & \mathbb{E}\left[\tau\left(Z\right)\mathds{1}\left\{ \tau\left(Z\right)>0\right\} \right]\\
 & = & \mu\left[1-\Phi\left(-\frac{\mu}{\sigma}\right)\right]+\sigma\phi\left(-\frac{\mu}{\sigma}\right)
\end{eqnarray*}
where $\Phi\left(\cdot\right)$ and $\phi\left(\cdot\right)$ are, respectively, the cdf and the probability density function (pdf) of the standard normal distribution. The target object is thus differentiable in the parameters of the data distribution. More generally, if the distribution of \(\tau(Z)\) has a density that is bounded in a neighborhood of zero, then it satisfies a standard margin bound with exponent one. Under suitable regularity conditions, such boundary control can restore the differentiability needed for conventional inference. In contrast, when no restriction is imposed on the distribution of \(\tau(Z)\) near zero, differentiability need not hold. We will see in Section \ref{sec_est} that the behavior of \(\tau(Z)\) near zero affects the smoothing bias and limiting variance, although our inference procedure does not impose a margin condition.

Since the $m\left(W,\boldsymbol{\gamma}\right)$ is non-differentiable in $\boldsymbol{\gamma},$ the results of \citet{Chernozhukov.et.al.2022a} cannot be directly applied. To be specific, $m\left(W,\boldsymbol{\gamma}\right)$ is not Gateaux differentiable at $\boldsymbol{\gamma}=\left(c,c\right)$ for $c\in\mathbb{R}.$ To see this, consider the Gateaux differential $dm\left(\left(c,c\right);\psi\right)$ of $m$ at $\left(c,c\right)$ in the direction $\psi=\left(\psi_{1},\psi_{2}\right)$ as follows.
\[
dm\left(\left(c,c\right);\psi\right)\equiv\lim_{\delta\rightarrow0}\frac{m\left(\left(c,c\right)+\delta\psi\right)-m\left(\left(c,c\right)\right)}{\delta}.
\]
If the limit exists for all directions $\psi,$ then $m$ is Gateaux differentiable at $\left(c,c\right).$ However, it is clear that the limit does not exist. Notice that for $\delta\neq0,$
\begin{eqnarray*}
\frac{m\left(\left(c,c\right)+\delta\psi\right)-m\left(\left(c,c\right)\right)}{\delta} & = & \frac{\delta\left(\psi_{1}-\psi_{2}\right)\mathds{1}\left\{ \delta\left(\psi_{1}-\psi_{2}\right)>0\right\} }{\delta}\\
 & = & \left(\psi_{1}-\psi_{2}\right)\mathds{1}\left\{ \delta\left(\psi_{1}-\psi_{2}\right)>0\right\} .
\end{eqnarray*}
Then,
\begin{eqnarray*}
\lim_{\delta\rightarrow0^{+}}\frac{m\left(\left(c,c\right)+\delta\psi\right)-m\left(\left(c,c\right)\right)}{\delta} & = & \lim_{\delta\rightarrow0^{+}}\left(\psi_{1}-\psi_{2}\right)\mathds{1}\left\{ \delta\left(\psi_{1}-\psi_{2}\right)>0\right\} \\
 & = & \left(\psi_{1}-\psi_{2}\right)\mathds{1}\left\{ \left(\psi_{1}-\psi_{2}\right)>0\right\} 
\end{eqnarray*}
and
\begin{eqnarray*}
\lim_{\delta\rightarrow0^{-}}\frac{m\left(\left(c,c\right)+\delta\psi\right)-m\left(\left(c,c\right)\right)}{\delta} & = & \lim_{\delta\rightarrow0^{-}}\left(\psi_{1}-\psi_{2}\right)\mathds{1}\left\{ \delta\left(\psi_{1}-\psi_{2}\right)>0\right\} \\
 & = & \left(\psi_{1}-\psi_{2}\right)\mathds{1}\left\{ \left(\psi_{1}-\psi_{2}\right)\leq0\right\} .
\end{eqnarray*}
The left and right limits are not the same, and hence $m\left(W,\boldsymbol{\gamma}\right)$ is not Gateaux differentiable at $\boldsymbol{\gamma}=\left(c,c\right)$ for $c\in\mathbb{R}.$

When $m\left(W,\boldsymbol{\gamma}\right)$ is nonlinear in $\boldsymbol{\gamma},$ \citet{Chernozhukov.et.al.2022a} linearizes the function and extends results for the linear case to the linearized function. The key assumption is that the remainder in the linearization, employing Gateaux differentiability in the neighborhood of the true parameter, is bounded by a constant. This assumption is crucial in deriving asymptotic normality of the DML estimator in the nonlinear case as it renders the remainder term negligible.

We introduce the sigmoid function to smooth the indicator function\footnote{The existence of Gateaux derivative of a functional is guaranteed as long as we stay in $C^{1}$ class. To be specific, consider a functional $\mathscr{F}:V\rightarrow\mathbb{R}$ is given as $\mathscr{F}\left(u\right)=\int_{\Omega}F\left(x,u\left(x\right),Du\left(x\right)\right)dx,\;\Omega\subseteq\mathbb{R}^{n},$ with functions $u:\Omega\rightarrow\mathbb{R}^{n}$ contained in some open subset $V$ of a function space $U\subseteq C^{1}\left(\Omega,\mathbb{R}^{n}\right).$ Under this specification, first variation (functional differential) of $\mathscr{F}$ exists and coincides with the Gateaux derivative of $\mathscr{F}.$ Thus, whenever $u$ and the integrand $F$ are of class $C^{1},$ the Gateaux derivative exists. In our setting, since the sigmoid function is of class $C^{1},$ the existence of Gateaux derivative of $m\left(W,\boldsymbol{\gamma}\right)$ is guaranteed.}. This smoothing function is characterized by a smoothing parameter which depends on the sample size. A notable feature is that the sigmoid function can be interpreted as the cumulative distribution function (cdf) of the logistic distribution, with the smoothing parameter scaling the distribution. Therefore, it is relatively convenient to derive an analytic expression for the approximation error introduced by smoothing using the analytic properties of the logistic distribution. This explicit analytic form facilitates the selection of an optimal smoothing parameter to control the trade-off between nuisance-estimation drift and smoothing approximation bias. Proper application of smoothing transforms a non-differentiable function into a differentiable one, while preserving the overall shape of the original function.

Despite the differentiability of the sigmoid function, the presence of the smoothing parameter means the results from \citet{Chernozhukov.et.al.2022a} do not directly carry over. \citet{Chernozhukov.et.al.2022a} assumes that the remainder term from linearization is bounded by a constant. When linearizing the sigmoid function, the remainder term depends on the smoothing parameter. This dependence causes the bound of the remainder term to increase as the sample size grows, thereby precluding straightforward application of the results of \citet{Chernozhukov.et.al.2022a}.

\section{Estimation and Inference}
\label{sec_est}

\begin{figure}[H]
\begin{centering}
\includegraphics[scale=0.38]{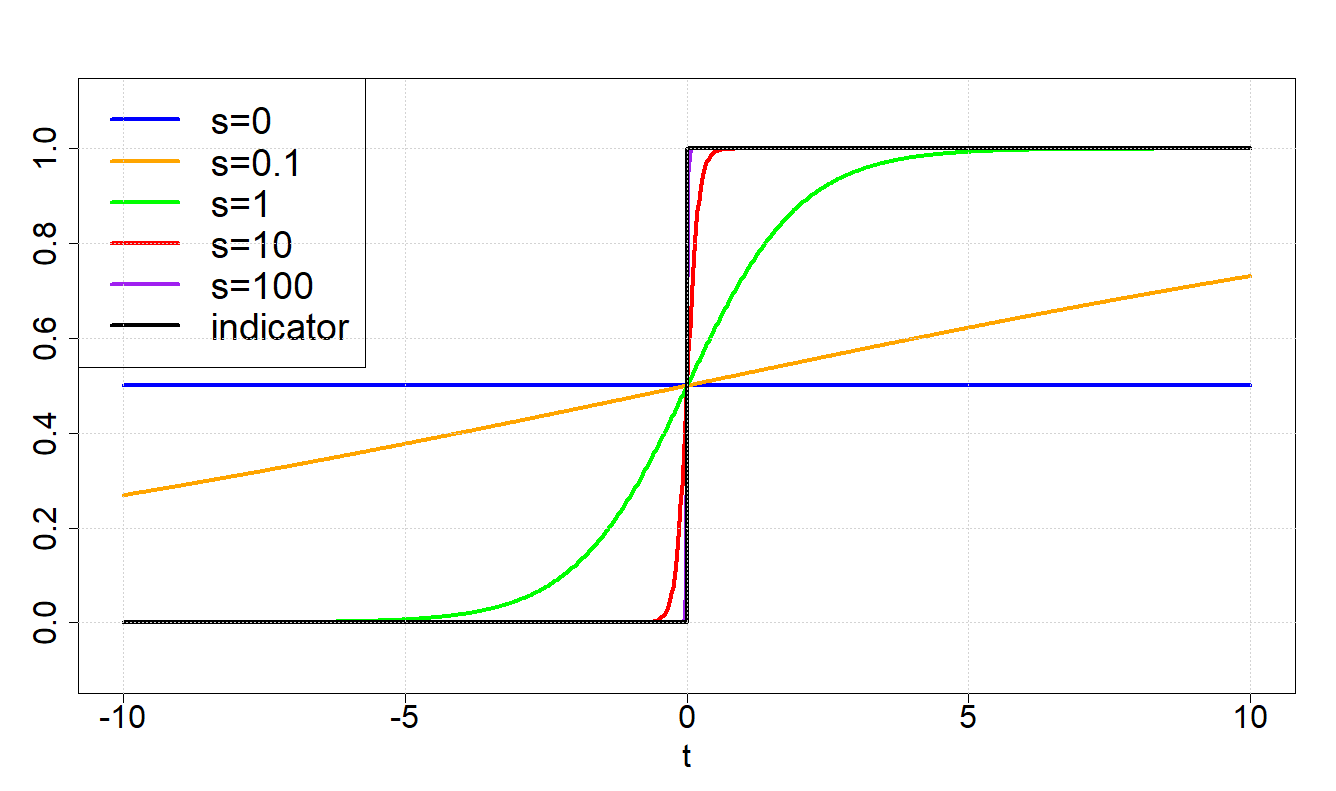}
\par\end{centering}
\noindent \caption{Sigmoid Function vs Indicator Function\label{fig:SigmoidvsIndicator}}
\end{figure}

\begin{figure}[H]
\begin{centering}
\includegraphics[scale=0.38]{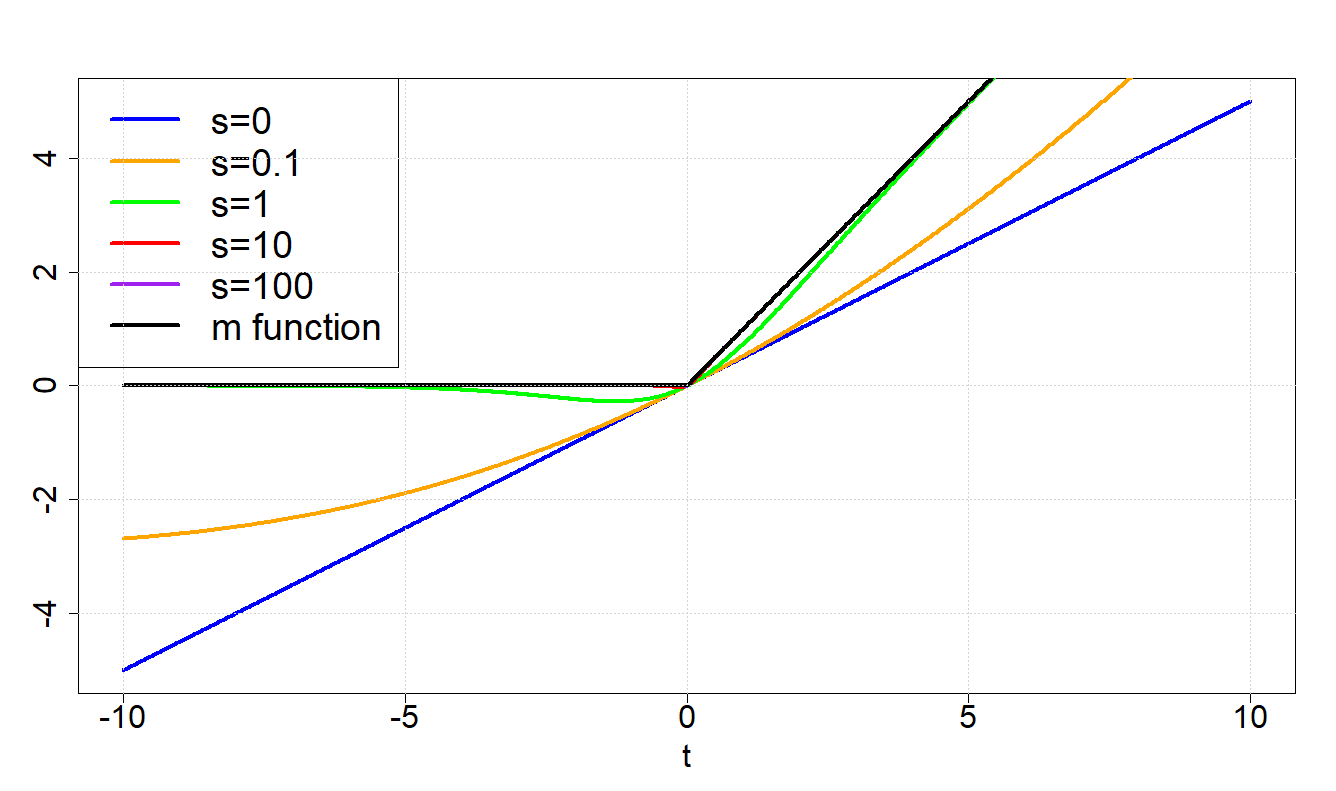}
\par\end{centering}
\noindent \caption{$m_{\mathrm{sig}}\left(W,\boldsymbol{\gamma}\right)$ vs $m\left(W,\boldsymbol{\gamma}\right)$\label{fig:mfunction}}
\end{figure}

The sigmoid function is defined as $f\left(t\right)=\frac{1}{1+\exp\left(-s_{n}t\right)}$ where $s_{n}>0$ can be interpreted as a smoothing parameter which depends on the sample size. As shown in Figure \ref{fig:SigmoidvsIndicator}, the sigmoid function approaches the indicator function as $s_{n}\rightarrow\infty.$
Let
\[
m_{\mathrm{sig}}\left(W,\boldsymbol{\gamma}\right)=\frac{\gamma_{1}\left(X\right)-\gamma_{2}\left(X\right)}{1+\exp\left(-s_{n}\left(\gamma_{1}\left(X\right)-\gamma_{2}\left(X\right)\right)\right)}
\]
be the smoothing function of
\[
m\left(W,\boldsymbol{\gamma}\right)=\left[\gamma_{1}\left(X\right)-\gamma_{2}\left(X\right)\right]\mathds{1}\left\{ \gamma_{1}\left(X\right)-\gamma_{2}\left(X\right)>0\right\} .
\]
Figure \ref{fig:mfunction} shows that $m_{\mathrm{sig}}\left(W,\boldsymbol{\gamma}\right)$ approaches $m\left(W,\boldsymbol{\gamma}\right)$ as $s_{n}\rightarrow\infty.$
Denote
\begin{eqnarray*}
\overline{\theta}_{\mathrm{sig}} & = & \mathbb{E}\left[m_{\mathrm{sig}}\left(W,\overline{\boldsymbol{\gamma}}\right)\right]\\
\overline{\theta} & = & \mathbb{E}\left[m\left(W,\overline{\boldsymbol{\gamma}}\right)\right].
\end{eqnarray*}
$\overline{\theta}$ can be viewed as the true parameter of interest, and $\overline{\theta}_{\mathrm{sig}}$ as a pseudo-true parameter.

\subsection{Estimators}
\label{subsec1}

As in \citet{Chernozhukov.et.al.2022a}, we can construct a DML estimator $\hat{\theta}_{\mathrm{sig}}.$ Consider the following decomposition for any $n\in\mathbb{N}.$
\begin{equation}
\sqrt{n}\left(\hat{\theta}_{\mathrm{sig}}-\overline{\theta}\right)=\sqrt{n}\left(\hat{\theta}_{\mathrm{sig}}-\overline{\theta}_{\mathrm{sig}}\right)+\sqrt{n}\left(\overline{\theta}_{\mathrm{sig}}-\overline{\theta}\right)\label{eq:asymptotic}
\end{equation}
Equation \eqref{eq:asymptotic} shows how the asymptotic distribution behaves when $\overline{\theta}$ is estimated by $\hat{\theta}_{\mathrm{sig}}.$ The first term on the right hand side $\sqrt{n}\left(\hat{\theta}_{\mathrm{sig}}-\overline{\theta}_{\mathrm{sig}}\right)$ is a random term which generates the asymptotic distribution of the estimator $\hat{\theta}_{\mathrm{sig}}$ around a pseudo-true parameter $\overline{\theta}_{\mathrm{sig}},$ and the second term $\sqrt{n}\left(\overline{\theta}_{\mathrm{sig}}-\overline{\theta}\right)$ is a nonrandom term which accounts for the error introduced by approximating the indicator function with the sigmoid function. As $\hat{\theta}_{\mathrm{sig}}$ and $\overline{\theta}_{\mathrm{sig}}$ involve $s_{n},$ both two terms are affected by the smoothing parameter $s_{n}$. The results of Theorem \ref{thm:oracle-bias-aware}, presented later in this section, make inference feasible when $s_{n}$ is chosen optimally.

Following equation \eqref{eq:asymptotic}, the DML estimator $\hat{\theta}_{\mathrm{sig}}$ is constructed as follows\footnote{In appendix, we briefly review DML where the parameter of interest depends linearly on a conditional expectation or nonlinearly on multiple conditional expectations}:
\begin{equation}
\hat{\theta}_{\mathrm{sig}}=\frac{1}{n}\sum_{\ell=1}^{L}\sum_{i\in I_{\ell}}\left\{ m_{\mathrm{sig}}\left(W_{i},\hat{\boldsymbol{\gamma}}_{\ell}\right)+\sum_{k=1}^{2}\hat{\alpha}_{k\ell}\left(X_{ki}\right)\left[Y_{ki}-\hat{\gamma}_{k\ell}\left(X_{ki}\right)\right]\right\} \label{eq:estimator}
\end{equation}
where the data $W_{i},$ $i=1,\cdots,n,$ are i.i.d., $I_{\ell},$ $\ell=1,\cdots L,$ is a partition of the observation index set $\left\{ 1,\cdots,n\right\} $ into $L$ distinct subsets of roughly equal size, and $\hat{\boldsymbol{\gamma}}_{\ell}=\left(\hat{\gamma}_{1\ell}\left(X_{1i}\right),\hat{\gamma}_{2\ell}\left(X_{2i}\right)\right)^{'}$ is the vector of regressions constructed by the observations not in $I_{\ell}.$ The estimator $\hat{\alpha}_{k\ell}\left(X_{ki}\right)$ of the Riesz representer specific to each regression is also constructed by the observations not in $I_{\ell}.$ Each $\hat{\alpha}_{k\ell}\left(X_{ki}\right)$ is obtained as follows. For each $k,$ denote $b_{k}\left(x_{k}\right)=\left(b_{k1}\left(x_{k}\right),\cdots,b_{kp}\left(x_{k}\right)\right)^{'}$ as a $p\times1$ dictionary vector specific to the $k$th regression $\gamma_{k}\left(x_{k}\right),$ and let $\hat{\boldsymbol{\gamma}}_{\ell,\ell^{'}}$ be the vector of regressions constructed by all observations not in either $I_{\ell}$ or $I_{\ell^{'}}.$ Also, let $\eta$ be a scalar, and $e_{k}$ be the $k$th column of the $2\times2$ identity matrix. Then, as in the equation (5.2) of \citet{Chernozhukov.et.al.2022a},
\begin{eqnarray}
\hat{\alpha}_{k\ell}\left(X_{ki}\right) & = & b_{k}\left(X_{ki}\right)^{'}\hat{\rho}_{k\ell}\nonumber \\
\hat{\rho}_{k\ell} & = & \underset{\rho}{\arg\min}\left\{ -2\hat{M}_{k\ell}^{'}\rho+\rho^{'}\hat{G}_{k\ell}\rho+2r_{k}\sum_{j=1}^{p}\left|\rho_{j}\right|\right\} \label{eq:estimator_alpha_rho}\\
\hat{M}_{k\ell} & = & \left(\hat{M}_{k\ell1},\cdots,\hat{M}_{k\ell p}\right)^{'}\nonumber \\
\hat{G}_{k\ell} & = & \frac{1}{n-n_{\ell}}\sum_{i\notin I_{\ell}}b_{k}\left(X_{ki}\right)b_{k}\left(X_{ki}\right)^{'}\nonumber \\
\hat{M}_{k\ell j} & = & \frac{d}{d\eta}\left(\frac{1}{n-n_{\ell}}\right)\sum_{\ell^{'}\neq\ell}\sum_{i\in I_{\ell^{'}}}m_{\mathrm{sig}}\left(W_{i},\hat{\gamma}_{\ell,\ell^{'}}+\eta e_{k}b_{kj}\right)\mid_{\eta=0},\;j=1,\cdots,p,\nonumber 
\end{eqnarray}
where $n_{\ell}$ is the number of observations in $I_{\ell},$ $b_{kj}$ is the $j$th element of the dictionary $b_{k}\left(x_{k}\right)$ as a function of $x_{k},$ and $r_{k}$ is the penalty size which must be chosen to be larger than $\sqrt{\ln\left(p\right)/n}.$

In the context of CATE estimation, this estimator can be categorized as a regularized T-learner, taking the difference between two conditional expectations and incorporating an additional debiasing correction term. Researchers may also consider alternative approaches. For example, \citet{Athey.and.Wager2019} treats the CATE function as a nuisance parameter and subsequently debiases it. It is generally difficult to argue that one method uniformly dominates another. \citet{Künzel.el.al.2019} provides comparisons of multiple learners and discusses the advantages of each method.

Our proposed estimator is an automatic DML for nonlinear functionals following \citet{Chernozhukov.et.al.2022a}. Its advantage over generic DML (e.g, \citet{Chernozhukov.et.al.2018}) is that it does not require prior knowledge of the explicit form of the correction term, that is, the Riesz representer. Moreover, even when a closed form is available, the generic DML approach, which first estimates the nuisance parameter such as the propensity score and then applies its analytical functional form, may not be optimal because of structural issues. For instance, to avoid numerical instability, \citet{Klosin2021} estimates continuous treatment effects using automatic DML, where the correction term is given as the multiplicative inverse of the (generalized) propensity score. In our setting, the correction term depends on the smoothing parameter. Automatic DML directly estimates the Riesz representer associated with the chosen value of that parameter rather than plugging estimated nuisance functions into its analytical form. Separately, the smoothing-parameter rule developed below balances upper bounds on nuisance-estimation drift and smoothing approximation bias.

Our analysis is based on the automatic DML using a sparse linear approximation of the Riesz representer as in \citet{Chernozhukov.et.al.2022a}. One may consider an adversarial approach to estimate the Riesz representer (\citet{Hirshberg.and.Wager.2021} and \citet{Chernozhukov.et.al.2024b}) within a broader functional class, but adversarial learning methods incur additional computational burdens. By introducing an approximate sparse specification of the Riesz representer, we can avoid these computational challenges by controlling the mean square approximation error and using Lasso. This approach also allows the identity of the important elements in the dictionary $b$ to remain unknown while still achieving the sparse approximation rate. Such a property is particularly useful for a policy maker who does not have prior knowledge of which elements of the dictionary are most relevant to the parameter of interest.

In contrast to the framework in \citet{Chernozhukov.et.al.2024a}, which considers a setting where the policy maker pre-selects a low-dimensional subset of covariates $Z_{\mathrm{sub}}$ (e.g., income level) from a high-dimensional set $Z$ and then identifies the CATE as
\begin{equation*}
    \tau\left(Z_{\mathrm{sub}}\right)=\mathbb{E}\left[\varphi_{0}\left(Y,Z,D\right)\mid Z_{\mathrm{sub}}\right]
\end{equation*}
with $\varphi_{0}\left(Y,Z,D\right)$ being the augmented inverse propensity weighted score (\citet{Robins.et.al1994}), and defines the value function as
\begin{equation*}
    V\left(\pi\right)\equiv\mathbb{E}\left[\pi\left(Z_{\mathrm{sub}}\right)\tau\left(Z_{\mathrm{sub}}\right)\right]=\mathbb{E}\left[\pi\left(Z_{\mathrm{sub}}\right)\varphi_{0}\left(Y,Z,D\right)\right]
\end{equation*}
so that the maximized average welfare gain is given by
\begin{equation*}
    V^{*}=\mathbb{E}\left[\mathds{1}\left(\tau\left(Z_{\mathrm{sub}}\right)\geq0\right)\varphi_{0}\left(Y,Z,D\right)\right],
\end{equation*}
our framework differs in an important way. In our approach, the maximized average welfare gain is defined as $\mathbb{E}\left[\mathds{1}\left(\tau\left(Z\right)\geq0\right)\tau\left(Z\right)\right]$ over the entire high-dimensional set $Z$, and the policy maker is not required to specify in advance which covariates are most relevant. The automatic DML procedure detects the relevant factors through estimation, thereby providing an alternative and flexible benchmark for policy evaluation.

\subsection{Theoretical Results}
\label{subsec2}

We impose the following regularity conditions (\citet{Chernozhukov.et.al.2022a}). For a matrix $A,$ define the norm $\left\Vert A\right\Vert _{1}=\sum_{i,j}\left|a_{ij}\right|.$ For a $p\times1$ vector $\rho,$ let $\rho_{J}$ be a $J\times1$ subvector of $\rho,$ and $\rho_{J^{c}}$ be the vector consisting of components of $\rho$ that are not in $\rho_{J}.$
\begin{assumption}
\label{assu:1}There exists $\frac{1}{4}<d_{\boldsymbol{\gamma}}<\frac{1}{2}$ such that $\left\Vert \hat{\gamma}_{k}-\overline{\gamma}_{k}\right\Vert =O_{p}\left(n^{-d_{\boldsymbol{\gamma}}}\right)$ for $k=1,2,$ where $\hat{\gamma}_{k}$ is a high-dimensional regression learner.
\end{assumption}
This assumption restricts the convergence rate of each $\hat{\gamma}_{k}.$ This is based on the results of \citet{Newey1994}, which shows that estimators which rely nonlinearly on unknown functions need to converge faster than $n^{-\frac{1}{4}}$ in terms of the norm.
\begin{assumption}
\label{assu:2}For each $k=1,2,$ $G_{k}=\mathbb{E}\left[b_{k}\left(X_{k}\right)b_{k}\left(X_{k}\right)^{'}\right]$ has the largest eigenvalue bounded uniformly in $n$ and there are $C,$ $c>0$ such that, for all $q\approx C\epsilon_{n}^{-2}$ with probability approaching 1,
\[
\min_{J\leq q}\min_{\left\Vert \rho_{J^{c}}\right\Vert _{1}\leq3\left\Vert \rho_{J}\right\Vert _{1}}\frac{\rho^{'}\hat{G}_{k}\rho}{\rho_{J}^{'}\rho_{J}}\geq c.
\]
\end{assumption}
This assumption is a sparse eigenvalue condition, which is generally assumed in Lasso literature (\citet{Bickel.et.al.2009}, \citet{Rudelson.and.Zhou2013}, and \citet{Belloni.and.Chernozhukov2013}).
\begin{assumption}
\label{assu:3}For each \(k=1,2\),
\[
\epsilon_n=o(r_k),
\qquad
r_k=o(n^c\epsilon_n)
\quad\text{for every }c>0,
\]
where \(\epsilon_n=n^{-d_{\boldsymbol{\gamma}}}\). In addition, there exists \(C>0\) such that \(p\leq Cn^C\).
\end{assumption}
This assumption characterizes the Lasso penalty size $r_{k},$ and restricts the growth rate of $p$ to be slower than some power of $n.$
\begin{assumption}
\label{assu:4}For some $\eta>0$, $\mathbb{E}\left[\left|Y_{k}-\overline{\gamma}_{k}\left(X_{k}\right)\right|^{2+\eta}\mid X_{k}\right]$ is bounded for $k=1,2.,$ and $\mathbb{E}\left|\overline{\tau}\left(X_{i}\right)\right|^{2+\eta}$ exists where $\overline{\tau}\left(X\right)=\overline{\gamma}_{1}\left(X\right)-\overline{\gamma}_{2}\left(X\right).$
\end{assumption}
This assumption imposes the finite moment conditions.
\begin{assumption}
\label{assu:5}For each k=1,2 and each cross-fitting fold $\ell,$
\begin{equation*}
    \left\Vert \hat{\alpha}_{k\ell}-\alpha_{k}\right\Vert =o_{p}\left(1\right),
\end{equation*}
and
\begin{equation*}
    \sqrt{n}\left\Vert \hat{\alpha}_{k\ell}-\alpha_{k}\right\Vert \left\Vert \hat{\gamma}_{k\ell}-\overline{\gamma}_{k}\right\Vert =o_{p}\left(1\right),
\end{equation*}
uniformly over $k$ and $\ell.$
\end{assumption}
This assumption is imposed as a high-level condition at the smoothing sequence used by the estimator, including the oracle choice
\(s_n=s_n^*\).
\begin{assumption}
\label{assu:6}$s_{n}\rightarrow\infty$ as $n\rightarrow\infty.$
\end{assumption}
This assumption characterizes the convergence rate of the smoothing parameter $s_{n}$.

The next proposition characterizes the asymptotic distribution of the estimator $\hat{\theta}_{\mathrm{sig}}$ around the pseudo-true parameter $\overline{\theta}_{\mathrm{sig}}.$
\begin{prop}
\label{prop:1}Let
\begin{eqnarray*}
\overline{\theta}_{\mathrm{sig}} & = & \mathbb{E}\left[m_{\mathrm{sig}}\left(W,\overline{\boldsymbol{\gamma}}\right)\right]\\
\psi_{\mathrm{sig}}\left(w\right) & = & m_{\mathrm{sig}}\left(W,\overline{\boldsymbol{\gamma}}\right)-\overline{\theta}_{\mathrm{sig}}+\sum_{k=1}^{2}\overline{\alpha}_{k}\left(x_{k}\right)\left[y_{k}-\overline{\gamma}_{k}\left(z_{k}\right)\right]\\
e\left(Z_{i}\right) & = & \mathrm{Pr}\left(D_{i}=1\mid Z_{i}\right)\\
\sigma_{1}^{2}\left(Z_{i}\right) & = & \mathrm{Var}\left(Y_{i}\mid D_{i}=1,Z_{i}\right)\\
\sigma_{2}^{2}\left(Z_{i}\right) & = & \mathrm{Var}\left(Y_{i}\mid D_{i}=0,Z_{i}\right)\\
\end{eqnarray*}
Under Assumptions \ref{assu:1}-\ref{assu:6},
\begin{equation}
    \sqrt{n}\left(\hat{\theta}_{\mathrm{sig}}-\overline{\theta}_{\mathrm{sig}}\right)=\frac{1}{\sqrt{n}}\sum_{i=1}^{n}\psi_{\mathrm{sig}}\left(W_{i}\right)+B_{n}+o_{p}\left(1\right)\label{eq:nonlinear}
\end{equation}
where
\[
\frac{1}{\sqrt{n}}\sum_{i=1}^{n}\psi_{\mathrm{sig}}\left(W_{i}\right)\overset{d}{\rightarrow}N\left(0,V\right),
\]
with
\begin{eqnarray*}
    V & = & \mathrm{Var}\left(\overline{\tau}\left(Z_{i}\right)\mathds{1}\left\{ \overline{\tau}\left(Z_{i}\right)>0\right\} \right)+\mathbb{E}\left[\mathds{1}\left\{ \overline{\tau}\left(Z_{i}\right)>0\right\} \left\{ \frac{\sigma_{1}^{2}\left(Z_{i}\right)}{e\left(Z_{i}\right)}+\frac{\sigma_{2}^{2}\left(Z_{i}\right)}{1-e\left(Z_{i}\right)}\right\} \right]\\
    & & +\frac{1}{4}\mathbb{E}\left[\mathds{1}\left\{ \overline{\tau}\left(Z_{i}\right)=0\right\} \left\{ \frac{\sigma_{1}^{2}\left(Z_{i}\right)}{e\left(Z_{i}\right)}+\frac{\sigma_{2}^{2}\left(Z_{i}\right)}{1-e\left(Z_{i}\right)}\right\} \right],
\end{eqnarray*}
and
\[
\left|B_{n}\right|\leq\frac{1}{2}\sqrt{n}s_{n}\left\Vert \hat{\gamma}-\overline{\gamma}\right\Vert ^{2}.
\]
\end{prop}
The following proposition characterizes $\overline{\theta}_{\mathrm{sig}}-\overline{\theta},$ which accounts for the approximation bias of our estimator.
\begin{prop}
\label{prop:2} The approximation error admits the following distribution-free bound
\[
0\leq\overline{\theta}-\overline{\theta}_{\mathrm{sig}}\leq\frac{W\left(e^{-1}\right)}{s_{n}},
\]
where $W\left(\cdot\right)$ is the Lambert $W$ function\footnote{$W\left(e^{-1}\right)$ is approximately 0.27846.}.
\end{prop}
$\quad$
\begin{rem}
Proposition \ref{prop:1} characterizes the asymptotic behavior of the DML estimator around the smoothed pseudo-true parameter \(\bar\theta_{\mathrm{sig}}\). The expansion consists of a Gaussian oracle term and a second-order nuisance linearization term \(B_{n}\). The Gaussian term has the usual \(\sqrt n\) normalization, but its limiting variance differs from the standard regular-boundary case when there is probability mass at the cutoff. In particular, since the derivative of the smoothed scalar map satisfies
\[
\frac{\partial}{\partial t}
\left\{
\frac{t}{1+\exp(-s_n t)}
\right\}
\longrightarrow
\mathds{1}\{t>0\}
+\frac{1}{2}\mathds{1}\{t=0\},
\]
the limiting variance contains an additional cutoff-mass component when $\mathrm{Pr}\left(\tau\left(Z\right)=0\right)>0$. This reflects the fact that units with zero CATE do not affect the level of the welfare functional, but they still affect the first-order residual correction through the limiting derivative of the smoothed score.

The variance formula in Proposition~\ref{prop:1} can be compared directly with the result of \citet{Chen.et.al.2025}. Let
\[
\Omega(Z)
=
\frac{\sigma_1^2(Z)}{e(Z)}
+
\frac{\sigma_2^2(Z)}{1-e(Z)}.
\]
In their unknown-covariate-distribution setting, their asymptotic variance for the welfare functional, after aligning notation, is
\[
\operatorname{Var}\!\left(
\bar\tau(Z)\mathds{1}\{\bar\tau(Z)>0\}
\right)
+
\mathbb{E}\!\left[
\mathds{1}\{\bar\tau(Z)>0\}\Omega(Z)
\right].
\]
Their regular-level-set condition, together with their continuous-covariate setting, implies \(\Pr(\bar\tau(Z)=0)=0\). Hence, when there is no probability mass at the cutoff, the variance \(V\) in Proposition~\ref{prop:1} reduces to their variance formula. In contrast, our framework allows \(\Pr(\bar\tau(Z)=0)>0\). Since the derivative of the smoothed moment converges to \(1/2\) at the cutoff, observations with zero CATE contribute the additional term
\[
\frac{1}{4}
\mathbb{E}\!\left[
\mathds{1}\{\bar\tau(Z)=0\}\Omega(Z)
\right]
\]
to the Gaussian variance component. This comparison concerns the Gaussian component of our asymptotic representation; our representation may additionally contain the non-negligible drift term \(B_n\) and the smoothing approximation bias.

The second term $B_{n}$ appears because the curvature of the smoothed moment function grows with the smoothing parameter \(s_n\). Proposition \ref{prop:1} shows that this term is bounded by \(\frac12\sqrt n s_n\|\widehat\gamma-\overline{\gamma}\|^2\), and hence the corresponding original-scale nuisance bias is bounded by \(\frac12s_n\|\widehat\gamma-\overline{\gamma}\|^2\). Therefore, unlike standard nonlinear DML results where the second-order remainder is asymptotically negligible, the present expansion keeps this term explicitly. This is also the key difference from nonsmoothed plug-in approaches under regular-boundary conditions, where the zero-CATE set has probability zero and the boundary movement term vanishes for the welfare functional.
\end{rem}
$\quad$
\begin{rem}
Proposition \ref{prop:2} characterizes the approximation bias from replacing the indicator function with the sigmoid function without imposing a margin condition. The sigmoid-smoothed estimand is weakly smaller than the original target, so the approximation bias is negative. This follows from the fact that the sigmoid lies below the indicator function on the positive side and above it on the negative side in a way that makes the smoothed welfare contribution smaller than the original welfare contribution.

The no-margin interpretation is important. Probability mass exactly at the cutoff does not itself create approximation bias. If \(\tau(Z)=0\), both the original moment and the smoothed moment are equal to zero. The bias instead comes from observations whose CATE values are close to, but not exactly equal to, zero. Without a margin condition, the distribution is allowed to place substantial nonzero mass arbitrarily close to the cutoff. Therefore, the approximation error admits the conservative uniform worst-case bound \(1/s_n\). Proposition \ref{prop:2} should be understood as a fully no-margin bound: it does not require $\mathrm{Pr}\left(\tau\left(Z\right)=0\right)=0$, and it also does not impose density control on the nonzero values of \(\tau(Z)\) near zero.
\end{rem}
$\quad$

Propositions~\ref{prop:1} and~\ref{prop:2}, together with equation~\eqref{eq:asymptotic}, reveal a trade-off between the second-order nuisance-estimation error and the smoothing approximation error. On the original scale, Proposition~\ref{prop:1} implies
\[
-\frac{1}{2}s_n
\left\|\widehat{\gamma}-\overline{\gamma}\right\|^2
\leq
\frac{B_{n}}{\sqrt{n}}
\leq
\frac{1}{2}s_n
\left\|\widehat{\gamma}-\overline{\gamma}\right\|^2,
\]
whereas Proposition~\ref{prop:2} determines both the magnitude and the direction of the smoothing approximation bias:
\[
-\frac{W(e^{-1})}{s_n}
\leq
\bar{\theta}_{\mathrm{sig}}-\bar{\theta}
\leq 0.
\]
Consequently, the leading non-Gaussian component of
\(\widehat{\theta}_{\mathrm{sig}}-\bar{\theta}\) satisfies
\[
-\left\{
\frac{1}{2}s_n
\left\|\widehat{\gamma}-\overline{\gamma}\right\|^2
+
\frac{W(e^{-1})}{s_n}
\right\}
\leq
\frac{B_{n}}{\sqrt{n}}
+
\bar{\theta}_{\mathrm{sig}}-\bar{\theta}
\leq
\frac{1}{2}s_n
\left\|\widehat{\gamma}-\overline{\gamma}\right\|^2.
\]
Thus, its worst-case absolute value is bounded by
\[
\frac{1}{2}s_n
\left\|\widehat{\gamma}-\overline{\gamma}\right\|^2
+
\frac{W(e^{-1})}{s_n}.
\]
Consider the following conservative upper-bound criterion for the leading asymptotic MSE:
\[
\frac{V}{n}
+
\left\{
\frac{1}{2}s_n
\left\|\widehat{\gamma}-\overline{\gamma}\right\|^2
+
\frac{W(e^{-1})}{s_n}
\right\}^2.
\]
Expanding this criterion gives
\[
\frac{V}{n}
+
\frac{1}{4}s_n^2
\left\|\widehat{\gamma}-\overline{\gamma}\right\|^4
+
W(e^{-1})
\left\|\widehat{\gamma}-\overline{\gamma}\right\|^2
+
\frac{W(e^{-1})^2}{s_n^2}.
\]
Because the variance term and the cross-product term do not depend on \(s_n\), minimizing the entire criterion is equivalent to minimizing
\[
\frac{1}{4}s_n^2
\left\|\widehat{\gamma}-\overline{\gamma}\right\|^4
+
\frac{W(e^{-1})^2}{s_n^2}.
\]
For \(\left\|\widehat{\gamma}-\overline{\gamma}\right\|>0\), the unique minimizer is
\[
s_n^*
=
\sqrt{2W(e^{-1})}
\left\|\widehat{\gamma}-\overline{\gamma}\right\|^{-1}.
\]
At this choice, the two components of the worst-case absolute-bias bound are equal:
\[
\frac{1}{2}s_n^*
\left\|\widehat{\gamma}-\overline{\gamma}\right\|^2
=
\frac{W(e^{-1})}{s_n^*}
=
\sqrt{\frac{W(e^{-1})}{2}}\,
\left\|\widehat{\gamma}-\overline{\gamma}\right\|.
\]
The minimized MSE upper-bound criterion is therefore
\[
\frac{V}{n}
+
2W(e^{-1})
\left\|\widehat{\gamma}-\overline{\gamma}\right\|^2.
\]
Under Assumption~\ref{assu:1},
\[
\left\|\widehat\gamma-\bar\gamma\right\|
=
O_p\left(n^{-d_{\boldsymbol{\gamma}}}\right),
\qquad
\frac{1}{4}<d_{\boldsymbol{\gamma}}<\frac{1}{2}.
\]
Therefore, the expansion and the minimized upper bound imply
\[
\widehat\theta_{\mathrm{sig}}-\bar\theta
=
O_p\left(n^{-d_{\boldsymbol{\gamma}}}\right).
\]
The centered influence-function component retains the usual \(n^{-1/2}\) order, but the worst-case convergence rate for the original unsmoothed target, as well as the order of the bias-aware confidence-interval half-length, is generally governed by the first-stage rate.

This differs from \citet{Whitehouse.et.al.2026}, which allows probability mass at exact treatment ties but impose a soft-margin condition on nonzero treatment gaps near the cutoff, making the smoothing bias asymptotically negligible at the \(n^{-1/2}\) scale. It also differs from \citet{Chen.et.al.2025}, whose parametric-rate result relies on a regular level set condition that excludes positive probability mass at the cutoff in their continuous-covariate setting. The slower worst-case rate obtained here reflects the weaker restriction that allows both probability mass at zero and unrestricted concentration of nonzero CATE values arbitrarily close to zero.

In what follows, \(\widehat\theta_{\mathrm{sig}}\), \(\bar\theta_{\mathrm{sig}}\), \(B_{n}\), and all score components are understood to be evaluated at \(s_n=s_n^*\). Define
$$
\widehat V_{\mathrm{sig}}
=
\frac{1}{n}
\sum_{\ell=1}^{L}
\sum_{i\in I_\ell}
\left[
m_{\mathrm{sig}}(W_i,\widehat\gamma_\ell)
+
\sum_{k=1}^{2}
\widehat\alpha_{k\ell}(X_{ki})
\left\{
Y_{ki}-\widehat\gamma_{k\ell}(X_{ki})
\right\}
-\widehat\theta_{\mathrm{sig}}
\right]^2,
$$
and
$$
\operatorname{se}(\widehat\theta_{\mathrm{sig}})
=
\sqrt{\frac{\widehat V_{\mathrm{sig}}}{n}}.
$$
Since \(s_n^*\) depends on the aggregate \(L_2\) error of the first-stage regression estimators, the smoothing parameter and confidence interval considered below are oracle quantities. The purpose of the following result is first to separate the asymptotically Gaussian component of the estimator from its potentially non-negligible drift and then to use an upper bound on that drift to construct a bias-aware confidence interval.

\begin{thm}
\label{thm:oracle-bias-aware}
Suppose that the conclusions of Propositions~\ref{prop:1} and~\ref{prop:2} hold at \(s_n=s_n^*\), with the Gaussian limit in Proposition~\ref{prop:1} also holding conditional on the first-stage estimates. Assume further that \(V\in(0,\infty)\) and
\[
\sqrt{n}\,
\operatorname{se}(\widehat\theta_{\mathrm{sig}})
\xrightarrow{p}\sqrt{V}.
\]
Then
$$
\frac{
\sqrt n
\bigl(\widehat\theta_{\mathrm{sig}}-\bar\theta\bigr)
-B_n
+\sqrt n
\bigl(\bar\theta-\bar\theta_{\mathrm{sig}}\bigr)
}{
\sqrt n\,
\operatorname{se}
\bigl(\widehat\theta_{\mathrm{sig}}\bigr)
}
\overset{d}{\longrightarrow}
N(0,1).
$$
Moreover, Propositions \ref{prop:1} and \ref{prop:2} imply that
$$
-\sqrt{2W(e^{-1})}
\|\widehat\gamma-\overline{\gamma}\|
\leq
\frac{B_n}{\sqrt n}
+
\bar\theta_{\mathrm{sig}}-\bar\theta
\leq
\sqrt{\frac{W(e^{-1})}{2}}
\|\widehat\gamma-\overline{\gamma}\|.
$$
Consequently,
$$
\left|
\frac{B_n}{\sqrt n}
+
\bar\theta_{\mathrm{sig}}-\bar\theta
\right|
\leq
\sqrt{2W(e^{-1})}
\|\widehat\gamma-\overline{\gamma}\|.
$$
\end{thm}

Theorem \ref{thm:oracle-bias-aware} does not assert that \(\sqrt n(\widehat\theta_{\mathrm{sig}}-\bar\theta)\) converges to a centered normal distribution. Instead, it separates the asymptotically Gaussian component from the potentially non-negligible drift. At \(s_n=s_n^*\), the upper bounds on the nuisance-estimation error and the smoothing approximation error are equal. The resulting drift bound is asymmetric because the smoothing approximation error is negative. For the confidence interval below, we conservatively use the larger absolute bound
$$
\sqrt{2W(e^{-1})}
\|\widehat\gamma-\overline{\gamma}\|.
$$
For \(b\geq 0\), let \(\operatorname{cv}_{1-\alpha}(b)\) denote the \(1-\alpha\) quantile of \(\lvert N(b,1)\rvert\). Conditional on the first-stage estimates, the Gaussian representation and absolute drift bound in Theorem~\ref{thm:oracle-bias-aware} place the estimator in the bounded-bias setting underlying the fixed-length confidence interval construction in Theorem 2.2 of \citet{Armstrong.and.Kolesar2020}. Applying that construction gives the following corollary.

\begin{cor}
\label{cor:oracle-ak}
Under the conditions of Theorem~\ref{thm:oracle-bias-aware}, define
$$
\mathrm{CI}^{\mathrm{AK,or}}_{1-\alpha}
=
\left[
\widehat\theta_{\mathrm{sig}}
\pm
\operatorname{se}(\widehat\theta_{\mathrm{sig}})
\operatorname{cv}_{1-\alpha}
\left(
\frac{
\sqrt{2W(e^{-1})}\,
\lVert\widehat\gamma-\overline{\gamma}\rVert
}{
\operatorname{se}(\widehat\theta_{\mathrm{sig}})
}
\right)
\right].
$$
Then, for every \(\alpha\in(0,1)\),
$$
\liminf_{n\to\infty}
\Pr_{P}
\left(
\bar\theta\in
\mathrm{CI}^{\mathrm{AK,or}}_{1-\alpha}
\right)
\geq 1-\alpha.
$$
\end{cor}

Corollary~\ref{cor:oracle-ak} converts the Gaussian representation in Theorem~\ref{thm:oracle-bias-aware} into a bias-aware confidence interval for \(\bar\theta\). The interval remains centered at \(\widehat\theta_{\mathrm{sig}}\), while the folded-normal critical value accounts for the maximum absolute drift. If the bias bound is small relative to the standard error, the critical value approaches the conventional two-sided normal critical value. If the bias is non-negligible, the critical value becomes correspondingly larger.

The interval is oracle because \(\|\widehat\gamma-\overline{\gamma}\|\) depends on the unknown regression functions. To construct a rate-calibrated version, let \(S_{\gamma,n}\) denote a positive normalization scale having the same units as the outcome. In the simulation and empirical implementations, this scale is the sample standard deviation of the outcome:
$$
S_{\gamma,n}
=
\widehat{\operatorname{sd}}(Y)
=
\left\{
\frac{1}{n-1}
\sum_{i=1}^{n}(Y_i-\bar Y)^2
\right\}^{1/2}.
$$
More generally, the theoretical analysis requires \(S_{\gamma,n}\) to converge in probability to a finite positive constant.

The role of \(S_{\gamma,n}\) is only to supply the appropriate outcome scale. It does not, by itself, estimate or bound the first-stage regression error. The magnitude of the maintained error radius is determined jointly by the dimensionless sensitivity parameter \(a_\gamma\), the scale \(S_{\gamma,n}\), and the convergence rate \(n^{-d_\gamma}\). Specifically, the implementation imposes the maintained bound
$$ \|\widehat\gamma-\overline{\gamma}\| \leq a_\gamma S_{\gamma,n}n^{-d_\gamma}. $$
Accordingly, the rate-calibrated smoothing parameter is
$$
\tilde{s}_{n}^{*}
=
\frac{\sqrt{2W(e^{-1})}}
{a_\gamma S_{\gamma,n}n^{-d_\gamma}},
$$
and the total absolute-bias bound used in the Armstrong-Kolesár critical value is
$$
\sqrt{2W(e^{-1})}
a_\gamma S_{\gamma,n}n^{-d_\gamma}.
$$
Because \(S_{\gamma,n}\) has outcome units, the maintained first-stage radius also has outcome units, while $\tilde{s}_{n}^{*}$ has inverse-outcome units. Thus,
$$
\tilde{s}_{n}^{*}
\{\widehat\gamma_1(X)-\widehat\gamma_2(X)\}
$$
is dimensionless, as required by the sigmoid function, and \(a_\gamma\) remains invariant to a rescaling of the outcome.

Suppose that, for some \(0\leq\delta<\alpha\), the maintained radius satisfies
\[
\limsup_{n\to\infty}
\Pr_P\left(
\left\|\widehat\gamma-\bar\gamma\right\|
>
a_\gamma S_{\gamma,n}n^{-d_\gamma}
\right)
\leq\delta.
\]
On the event that the radius condition holds, Propositions \ref{prop:1} and \ref{prop:2} bound the absolute drift by
\[
\sqrt{2W(e^{-1})}\,
a_\gamma S_{\gamma,n}n^{-d_\gamma}.
\]
Provided that the Gaussian approximation and standard-error consistency in Theorem~\ref{thm:oracle-bias-aware} hold at \(s_n=\widetilde{s}_n^*\), applying the Armstrong-Kolesár critical value at level \(1-\alpha+\delta\) and accounting for the radius-failure probability yields pointwise asymptotic coverage of at least \(1-\alpha\). The particular choices of \(a_\gamma\), \(d_\gamma\), and the probability attached to the maintained radius are discussed separately in the simulation and empirical sections.

\begin{rem}
Although our asymptotic analysis considers \(s_n\to\infty\), the opposite limit provides a useful interpretation of the smoothing parameter. In particular,
\[
\lim_{s_n\to 0}m_{\mathrm{sig}}(W,\gamma)
=
\frac{1}{2}\{\gamma_1(X)-\gamma_2(X)\},
\]
so the corresponding population estimand converges to one half of the ATE. Moreover, the derivatives of the smoothed moment with respect to \(\gamma_1\) and \(\gamma_2\) converge to \(1/2\) and \(-1/2\), respectively. Consequently, the associated Riesz representers and orthogonal score converge to those for one half of the ATE. The moment is linear in this limiting case, so the automatic DML construction for linear functionals in \citet{Chernozhukov.et.al.2022a} applies. Thus, the sigmoid construction connects one half of the ATE as \(s_n\to0\) with the maximized average welfare gain as \(s_n\to\infty\).
\end{rem}

\subsection{Alternative Smoothing Function}
\label{subsec3}

Our target parameter is defined as the expectation of the moment function
\[
m\left(W,\boldsymbol{\gamma}\right)=\left(\gamma_{1}\left(X\right)-\gamma_{2}\left(X\right)\right)\mathds{1}\left\{ \gamma_{1}\left(X\right)-\gamma_{2}\left(X\right)>0\right\} 
\]
where the moment function is equivalent to $\max\left\{ \gamma_{1}\left(X\right)-\gamma_{2}\left(X\right),0\right\} $. In our approach, we smooth the indicator function by using a sigmoid function, thereby obtaining the smoothed moment function
\[
m_{\mathrm{sig}}\left(W,\boldsymbol{\gamma}\right)\equiv\frac{\gamma_{1}\left(X\right)-\gamma_{2}\left(X\right)}{1+\exp\left(-s_{n}\left(\gamma_{1}\left(X\right)-\gamma_{2}\left(X\right)\right)\right)}.
\]
Alternatively, one may smooth the maximum directly via the log-sum-exp (LSE) function (\citet{Levis.et.al.2023}). In that case the smoothing function is defined as
\[
m_{\mathrm{LSE}}\left(W,\boldsymbol{\gamma}\right)\equiv\frac{1}{h_{n}}\log\left(1+\exp\left(h_{n}\left(\gamma_{1}\left(X\right)-\gamma_{2}\left(X\right)\right)\right)\right),
\]
where $h_{n}$ plays the same role as the smoothing parameter in our approach. Notably, the Gateaux derivative of $m_{\mathrm{LSE}}\left(W,\boldsymbol{\gamma}\right)$ in the direction of the true treatment effect difference is exactly $m_{\mathrm{sig}}\left(W,\boldsymbol{\gamma}\right)$. Therefore, when assessing the approximation error introduced by smoothing, both the sigmoid‐based and LSE‐based approaches are fundamentally linked to the logistic distribution and exhibit equivalent theoretical properties.

On the other hand, one can observe that
\[
m_{\mathrm{sig}}\left(W,\boldsymbol{\gamma}\right)\leq m\left(W,\boldsymbol{\gamma}\right)\leq m_{\mathrm{LSE}}\left(W,\boldsymbol{\gamma}\right),
\]
where the first inequality holds with equality at the cutoff, whereas the second inequality is strict for finite \(h_n\). In particular,
\[
m_{\mathrm{LSE}}(W,\boldsymbol{\gamma})
=
\frac{\log 2}{h_n}
\]
at the cutoff. As a result, estimates based on the sigmoid smoothing are likely to be smaller than those based on the LSE smoothing. Which smoothing method to adopt can ultimately depend on the policy maker's preference. For example, if one wishes to avoid overestimating the welfare gain, a conservative policy maker may choose the sigmoid function.

\section{Simulation}
\label{sec_sim}

This section examines the finite-sample performance of the estimator and the bias-aware confidence interval. The data-generating process assigns positive probability to a zero conditional average treatment effect, so the simulation does not rely on a margin condition that excludes observations at the treatment cutoff. We compare different values of $a_\gamma$ in terms of bias, estimated standard error, root mean squared error, and the coverage and length of the Armstrong-Kolesár confidence interval.

\subsection{Simulation design}

For each observation, let
\[
Z_i=(G_i,X_{i1},\ldots,X_{i,100})',
\]
where $G_i\sim\operatorname{Bernoulli}(0.75)$, $X_{i1}\sim\operatorname{Unif}[-1,1]$, and $X_{ij}\sim N(0,1)$ for $j=2,\ldots,100$. These variables are mutually independent. Treatment is assigned according to $D_i\sim\operatorname{Bernoulli}(0.5)$, independently of the covariates, and the observed outcome is generated as
\[
Y_i=0.5X_{i2}-0.5X_{i3}+2D_iG_iX_{i1}+\varepsilon_i,
\qquad \varepsilon_i\sim N(0,1),
\]
where the noise is independent of treatment and the covariates. Let
\[
\gamma_1(z)=\mathbb{E}[Y_i\mid D_i=1,Z_i=z],
\qquad
\gamma_2(z)=\mathbb{E}[Y_i\mid D_i=0,Z_i=z].
\]
The conditional average treatment effect is
\[
\tau(Z_i)=\gamma_1(Z_i)-\gamma_2(Z_i)=2G_iX_{i1}.
\]
Consequently,
\[
\Pr\{\tau(Z_i)=0\}=0.25,
\]
and the true target parameter is
\[
\bar\theta=\mathbb{E}\bigl[\tau(Z_i)\mathds{1}\{\tau(Z_i)>0\}\bigr]=0.375.
\]
We use one dictionary specification throughout the simulation. It consists of an intercept, $G_i$, the 100 primitive covariates, and ten interaction terms between $G_i$ and the Legendre polynomials of $X_{i1}$:
\[
G_iP_j(X_{i1}),\qquad j=1,\ldots,10,
\]
where $P_j(\cdot)$ denotes the Legendre polynomial of degree $j$. The resulting dictionary has dimension $p=112$. The same dictionary is used to estimate the conditional mean functions and the Riesz representers. Non-intercept terms are centered and scaled using the full simulated sample within each replication. We employ five-fold cross-fitting, with nested cross-fitting within the outer training samples to construct the derivatives entering the Riesz regressions.

Following the loading-adjusted Lasso procedure of \citet{Chernozhukov.et.al.2022a}, the penalty degree for each conditional mean regression is
\[
\lambda_{\gamma,k\ell}
=c_\gamma\frac{\Phi^{-1}\{1-0.1/(2p)\}}{\sqrt{n_{k\ell}}},
\qquad c_\gamma=1,
\]
where $n_{k\ell}$ is the relevant arm-specific training-sample size. For the Riesz regressions, we use
\[
r_{\alpha,\ell}
=c_\alpha N_\ell^{-d_\gamma}\log N_\ell,
\qquad c_\alpha=1,\qquad d_\gamma=0.49,
\]
where $N_\ell=n-|I_\ell|$ is the size of the entire outer training sample, including both treatment arms, and $\log$ denotes the natural logarithm. This penalty degree is used for both Riesz regressions, together with their respective penalty loadings. The multiplier $c_\alpha=1$ is fixed as a unit benchmark throughout the simulation.

The sample size is $n=2000$, and the results are based on 1000 Monte Carlo replications. In each replication, the outcome scale is estimated by
\[
S_{\gamma,n}
=\left\{\frac{1}{n-1}\sum_{i=1}^n(Y_i-\bar Y)^2\right\}^{1/2}.
\]
We construct the rate-based radius and corresponding smoothing parameter as
\[
R_{\gamma,n}=a_\gamma S_{\gamma,n}n^{-d_\gamma},
\qquad
s_n^*=\frac{\sqrt{2W(e^{-1})}}{R_{\gamma,n}},
\]
and consider
\[
a_\gamma\in\{1,10,12\}.
\]
The value $a_\gamma=1$ provides a unit-normalized benchmark, while $a_\gamma=10$ provides an intermediate sensitivity specification. The value $a_\gamma=12$ is the pilot-calibrated specification targeting a radius-coverage probability of $0.975$ for this DGP. Because $s_n^*$ is inversely related to $a_\gamma$, increasing $a_\gamma$ results in a smaller smoothing parameter and a greater degree of smoothing.

We construct the Armstrong-Kolesár interval using the absolute drift bound in Corollary \ref{cor:oracle-ak}, replacing the unknown first-stage error norm by $R_{\gamma,n}$. For a common comparison across specifications, all intervals use the folded-normal critical value at level $0.975$. This choice targets pointwise asymptotic coverage of at least $0.95$ when the radius-failure probability is at most $0.025$ and the remaining conditions for the Gaussian approximation and standard-error consistency hold. The smaller values of $a_\gamma$ are included as sensitivity specifications; their coverage performance is evaluated separately from whether the proposed radius bounds the first-stage error. We also report the coverage of the conventional interval $\widehat\theta_{\mathrm{sig}}\pm1.96\, \widehat{\mathrm{se}}(\widehat\theta_{\mathrm{sig}})$.

\subsection{Results}

Table~\ref{tab:simulation-results} reports the Monte Carlo results. Bias and RMSE are calculated relative to the original target $\bar\theta=0.375$. Mean SE is the average estimated standard error across replications. Normal and AK coverage are the fractions of replications in which the corresponding intervals contain the original target.

\begin{table}[htbp]
\centering
\small
\setlength{\tabcolsep}{4pt}
\begin{tabular}{rrrrrrr}
\hline
$a_\gamma$ & Bias & Mean SE & RMSE
& Normal coverage & AK coverage & Mean AK length \\
\hline
1  & $-0.0210$ & 0.0275 & 0.0358 & 0.842 & 0.971 & 0.1590 \\
10 & $-0.0841$ & 0.0281 & 0.0891 & 0.173 & 1.000 & 0.6193 \\
12 & $-0.1049$ & 0.0279 & 0.1089 & 0.049 & 1.000 & 0.7204 \\
\hline
\end{tabular}
\caption{Monte Carlo performance across values of $a_\gamma$}
\label{tab:simulation-results}
\end{table}

The unit-normalized specification $a_\gamma=1$ produces the smallest absolute bias and RMSE among the three specifications. Its bias is $-0.0210$, while its mean estimated standard error is $0.0275$. As $a_\gamma$ increases to 10 and 12, the bias becomes more negative, reaching $-0.0841$ and $-0.1049$, respectively. The corresponding RMSE increases from $0.0358$ to $0.0891$ and $0.1089$. In contrast, the mean estimated standard error remains between $0.0275$ and $0.0281$. The Monte Carlo standard deviations of the estimator range from $0.0290$ to $0.0295$, so the estimated standard errors are approximately five percent smaller than the observed sampling standard deviations. The differences in RMSE across specifications are therefore primarily associated with changes in bias.

The increasing downward shift reflects stronger smoothing, but smoothing does not account for the entire finite-sample bias. At $a_\gamma=12$, the average population value of the smoothed functional is $0.2957$, compared with the original target of $0.375$. The average feasible estimate is $0.2701$, which lies a further $0.0256$ below the smoothed population target. An oracle diagnostic that replaces the estimated Riesz representers by their true counterparts, while holding the conditional mean estimates and smoothing parameter fixed, raises the average estimate to $0.2971$. At $a_\gamma=1$, the corresponding smoothed population target is $0.3744$, whereas the average feasible estimate is $0.3540$. These diagnostics show that nuisance estimation contributes to the observed downward bias in addition to smoothing.

The conventional normal interval has coverage of $0.842$, $0.173$, and $0.049$ as $a_\gamma$ increases from 1 to 10 and 12. By comparison, the Armstrong-Kolesár interval has coverage of $0.971$, $1.000$, and $1.000$. The Monte Carlo standard error of the coverage estimate $0.971$ is approximately $0.0053$; coverage of $1.000$ means that all 1,000 intervals in the simulation contain the target. The higher coverage under the larger radius specifications is accompanied by substantially longer intervals: the mean AK length increases from $0.1590$ to $0.6193$ and $0.7204$. Thus, these specifications provide conservative coverage in this design, with a substantial cost in interval length.

To assess the radius calibration, we evaluate the fold-specific conditional mean estimates against the true conditional means on an independent sample of 10,000 covariate draws in each replication. Using the resulting approximation to the aggregate two-arm $L_2$ error, the radius covers the first-stage error in $0.000$, $0.545$, and $0.974$ of the replications for $a_\gamma=1$, 10, and 12, respectively. The last proportion is close to the target probability of $0.975$, with a Monte Carlo standard error of approximately $0.0050$. The smaller radius specifications do not achieve the intended radius coverage at this sample size, even though their AK intervals have coverage above $0.95$ in this experiment. Their observed interval coverage should therefore be interpreted as finite-sample sensitivity evidence rather than as verification of the radius condition.

\begin{figure}[!htbp]
\centering
\includegraphics[width=0.92\textwidth]{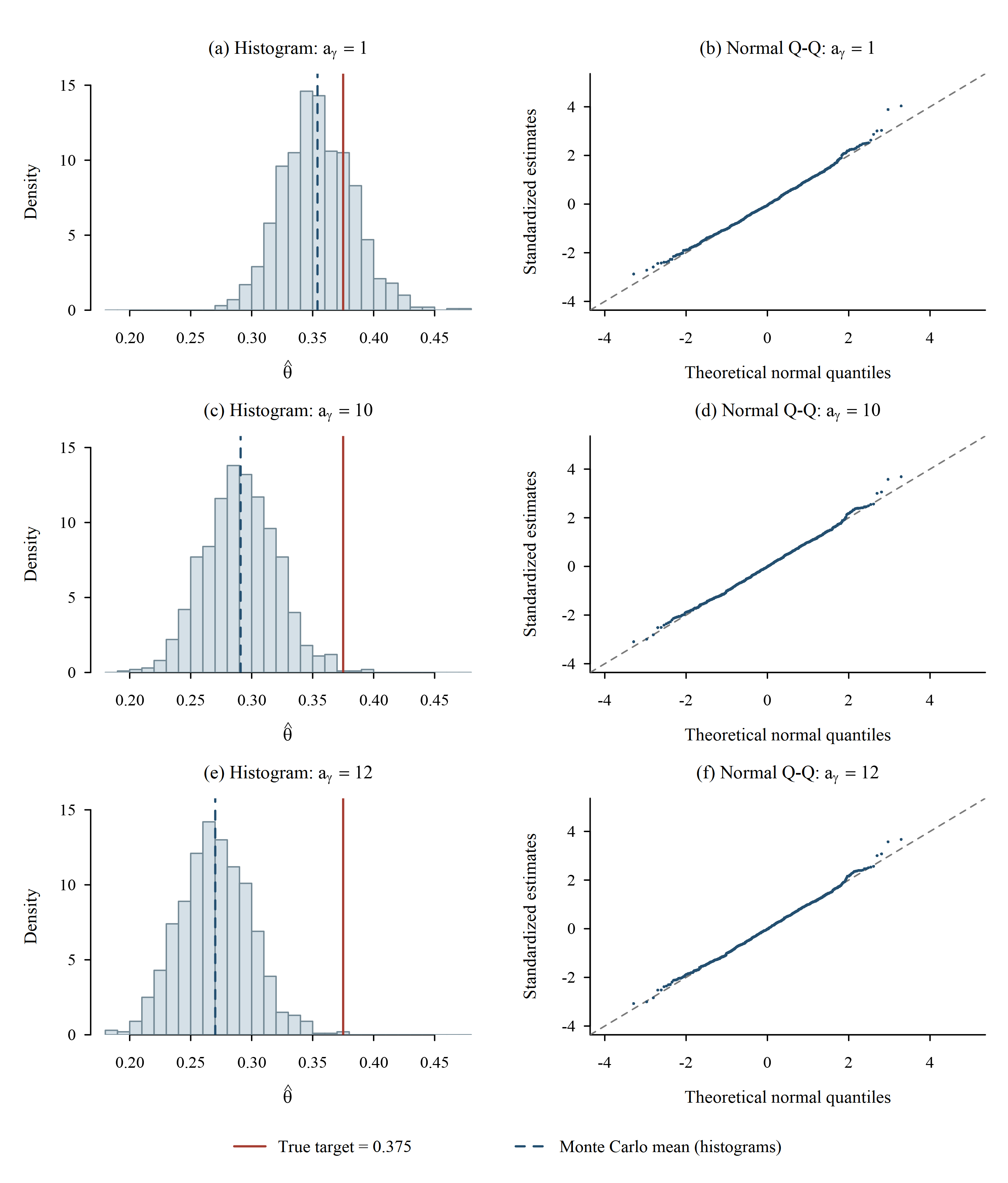}
\caption{Sampling distributions and normal Q--Q plots}
\label{fig:simulation-distributions}
\medskip
\begin{minipage}{0.95\textwidth}
\footnotesize
\emph{Notes:} Rows correspond to $a_\gamma=1$, 10, and 12, respectively. Each panel uses 1,000 replications with $n=2{,}000$ and $c_\alpha=1$. In the histograms, the solid red line marks the true target $\bar\theta=0.375$, and the dashed blue line marks the Monte Carlo mean. The Q-Q plots use estimates centered at their setting-specific Monte Carlo mean and divided by their Monte Carlo sample standard deviation. The dashed gray line is the identity reference.
\end{minipage}
\end{figure}

Figure~\ref{fig:simulation-distributions} presents the histograms and normal Q-Q plots for the three specifications. The histograms use common axes and display both the true target and the Monte Carlo mean. All three distributions are centered below the target, with a larger downward shift for larger $a_\gamma$. For the Q-Q plots, each estimator is centered at its own Monte Carlo mean and divided by its Monte Carlo standard deviation. The plots are approximately linear over most of their ranges, with some deviations in the extreme tails. They indicate that an approximately Gaussian distributional shape can coexist with substantial drift relative to the original target; they do not establish a normal approximation centered at that target.

\section{Empirical Analysis}
\label{sec_emp}

We apply our method to experimental data from the National Job Training Partnership Act (JTPA) Study, following the empirical setting of \citet{Kitagawa.and.Tetenov2018}. The sample consists of 9,223 observations. The outcome variable is earnings, and the binary treatment variable indicates assignment to a job training program. The five base covariates are age, education, indicators for Black and Hispanic individuals, and earnings in the year preceding treatment assignment. The parameter $\bar\theta$ measures the average earnings gain, relative to assigning no one to treatment, from treating individuals with a positive conditional average treatment effect. While \citet{Kitagawa.and.Tetenov2018} use education and pre-treatment earnings in their empirical welfare maximization rules, we include all five covariates in the conditional mean specifications.

We consider four dictionary specifications. Specification (1) includes an intercept, the five base covariates, squared terms for age, education, and pre-treatment earnings, and pairwise interactions among the five base covariates, excluding the interaction between the Black and Hispanic indicators. Its dimension is 18. Specification (2) includes an intercept, the five base covariates, and squared and cubic terms for age, education, and pre-treatment earnings. Its dimension is 12. Specification (3) extends Specification (2) by adding fourth-order terms for age and education, resulting in a dictionary of dimension 14. Specification (4) further adds the pairwise interactions used in Specification (1) and the fifth- and sixth-order terms for age and education. Its dimension is 27.

Non-intercept dictionary terms are centered and scaled using the full sample. Earnings are demeaned but remain in their original dollar units. We use five-fold cross-fitting with folds stratified by treatment status, and the same loading-adjusted Lasso procedure as in the simulation. The folds are held fixed across dictionary and sensitivity specifications. Nested cross-fitting within each outer training sample is used to construct the derivatives entering the Riesz regressions. We set $c_\gamma=c_\alpha=1$ and use the Riesz penalty degree
\[
r_{\alpha,\ell}
=c_\alpha N_\ell^{-d_\gamma}\log N_\ell,
\qquad d_\gamma=0.49,
\]
where $N_\ell=n-|I_\ell|$ is the size of the entire outer training sample, including both treatment arms. All reported nuisance fits satisfy the loading and coordinate-convergence criteria.

We set $S_{\gamma,n}$ equal to the sample standard deviation of earnings, $17,070.81$, and construct the first-stage radius and smoothing parameter as
\[
R_{\gamma,n}=a_\gamma S_{\gamma,n}n^{-d_\gamma},
\qquad
s_n^*=\frac{\sqrt{2W(e^{-1})}}{R_{\gamma,n}}.
\]
We report results for $a_\gamma\in\{1,3,6,12\}$. These values span progressively larger maintained first-stage error radii and therefore imply progressively smaller smoothing parameters. Because the true $L_2$ error of the conditional mean estimators is unknown, all four values are treated as sensitivity specifications. The value $a_\gamma=1$ is a unit-normalized benchmark, rather than an empirically calibrated radius for the JTPA data.

For each sensitivity specification, we maintain that the first-stage radius condition fails with probability at most $0.025$. The intervals use the Armstrong-Kolesár folded-normal critical value at level $0.975$. Accounting for the radius-failure probability, the resulting intervals have pointwise asymptotic coverage of at least $0.95$ under the maintained radius condition and the remaining conditions for the Gaussian approximation and standard-error consistency.

\begin{table}[!htbp]
\centering
\small
\setlength{\tabcolsep}{7pt}
\begin{tabular}{crrrr}
\hline
Specification & $a_\gamma$ & Estimate & SE & 95\% AK CI \\
\hline
(1) & 1  & 1,179 & 323 & $(391,\,1{,}968)$ \\
(1) & 3  & 1,011 & 311 & $(-34,\,2{,}056)$ \\
(1) & 6  & 860   & 267 & $(-534,\,2{,}254)$ \\
(1) & 12 & 748   & 224 & $(-1{,}436,\,2{,}932)$ \\
\hline
(2) & 1  & 1,200 & 320 & $(416,\,1{,}984)$ \\
(2) & 3  & 1,029 & 311 & $(-16,\,2{,}074)$ \\
(2) & 6  & 869   & 267 & $(-526,\,2{,}263)$ \\
(2) & 12 & 748   & 224 & $(-1{,}436,\,2{,}931)$ \\
\hline
(3) & 1  & 1,208 & 322 & $(420,\,1{,}995)$ \\
(3) & 3  & 1,033 & 311 & $(-14,\,2{,}079)$ \\
(3) & 6  & 864   & 267 & $(-531,\,2{,}258)$ \\
(3) & 12 & 745   & 224 & $(-1{,}438,\,2{,}929)$ \\
\hline
(4) & 1  & 1,193 & 326 & $(397,\,1{,}989)$ \\
(4) & 3  & 1,016 & 312 & $(-32,\,2{,}064)$ \\
(4) & 6  & 856   & 266 & $(-539,\,2{,}250)$ \\
(4) & 12 & 742   & 224 & $(-1{,}441,\,2{,}925)$ \\
\hline
\end{tabular}
\caption{Estimation results for the JTPA application}
\label{tab:jtpa-results}
\end{table}

Table~\ref{tab:jtpa-results} reports the estimates, standard errors, and Armstrong-Kolesár confidence intervals. All monetary quantities are expressed in dollars and rounded to the nearest dollar. Conditional on a given value of $a_\gamma$, both the estimates and standard errors are stable across the four dictionary specifications. At the unit-normalized benchmark $a_\gamma=1$, the estimated welfare gain ranges from 1,179 to 1,208 dollars, with standard errors between 320 and 326 dollars. The lower endpoints of the four confidence intervals range from 391 to 420 dollars, so all four intervals exclude zero.

As $a_\gamma$ increases, the corresponding smoothing parameter decreases and the estimated welfare gain declines. For $a_\gamma=3$, the estimates range from 1,011 to 1,033 dollars, and the lower endpoints of the confidence intervals range from $-34$ to $-14$ dollars. For $a_\gamma=6$, the estimates range from 856 to 869 dollars, with lower endpoints between $-539$ and $-526$ dollars. For $a_\gamma=12$, the estimates range from 742 to 748 dollars, with lower endpoints between $-1{,}441$ and $-1{,}436$ dollars. Thus, the intervals include zero under each of the larger-radius sensitivity specifications, although the point estimates remain positive.

The estimated standard errors decrease as $a_\gamma$ increases, but the larger allowance for nuisance-estimation and smoothing bias causes the Armstrong-Kolesár intervals to widen. The total absolute bias allowance is approximately 145, 436, 872, and 1,744 dollars for $a_\gamma=1,3,6,12$, respectively. Consequently, the differences across values of $a_\gamma$ are considerably larger than the differences across dictionary specifications. The empirical conclusion about whether the welfare gain is statistically positive is therefore sensitive to the maintained first-stage error radius: the intervals exclude zero under the unit-normalized benchmark, but include zero under the larger-radius specifications.

For comparison, \citet{Kitagawa.and.Tetenov2018} reports an estimated welfare gain of 1,340 dollars with a 95 percent confidence interval of $(441,2{,}239)$ for the EWM quadrant rule, 1,364 dollars with an interval of $(398,2{,}330)$ for the EWM linear rule, and 1,489 dollars with an interval of $(374,2{,}603)$ for the EWM linear rule including squared and cubic terms for education. Across all dictionary specifications and values of $a_\gamma$, our point estimates are smaller than those reported for the EWM rules. This pattern is qualitatively consistent with the downward approximation bias induced by sigmoid smoothing. The corresponding confidence intervals overlap substantially with their reported intervals.

\section{Conclusion}
\label{sec_conc}

This paper focuses on debiased machine learning when nuisance parameters appear in indicator functions and there is a high-dimensional vector of covariates. We propose a DML estimator where the indicator function is smoothed. The asymptotic theory motivates an oracle smoothing parameter that balances upper bounds on nuisance-estimation drift and smoothing approximation bias, together with a bias-aware confidence interval that accounts for the remaining drift. Simulations and empirical exercise corroborate these results.

There are several ways in which the proposed procedure could be developed further. First, it would be useful to extend the specific bias-aware DML approach developed here to other non-differentiable functionals and alternative smoothing methods. Second, the bias-aware confidence intervals can be conservative because they rely on a worst-case bound on the total bias. Deriving sharper bias bounds could therefore improve precision. Third, developing a fully data-driven procedure for calibrating the first-stage error radius while preserving the coverage guarantee remains an important direction for future research.

\appendix

\section{DML and Orthogonal Moment Functions}
\label{app_dml}

This section reviews DML where the parameter of interest depends linearly
on a conditional expectation or nonlinearly on multiple conditional
expectations, which are developed in later sections. Notations generally
follow \citet{Chernozhukov.et.al.2022a}. Let $W=\left(Y,X^{'}\right)^{'}$
denote an observation where $Y$ is an outcome variable with a finite
second moment and $X$ is a high-dimensional vector of covariates.
Let
\[
\boldsymbol{\gamma}_{0}\left(x\right)\equiv\mathbb{E}\left[Y\mid X=x\right]
\]
be the conditional expectation of $Y$ given $X\in\mathcal{X}.$ Let
$\boldsymbol{\gamma}:\mathcal{X}\rightarrow\mathbb{R}$ be a function
of $X.$ Define $m\left(w,\boldsymbol{\gamma}\right)$ as a function
of the function $\boldsymbol{\gamma}$ (i.e. a functional of $\boldsymbol{\gamma}$),
which depends on an observation $w.$ The parameter of interest $\theta_{0}$
has the following expression:
\[
\theta_{0}=\mathbb{E}\left[m\left(W,\boldsymbol{\gamma}_{0}\right)\right].
\]
\citet{Chernozhukov.et.al.2022a} present examples where $m\left(W,\boldsymbol{\gamma}\right)$
is linear and nonlinear in $\boldsymbol{\gamma}.$ The examples where
it is linear in $\boldsymbol{\gamma}$ are the average policy effect,
weighted average derivative, average treatment effect and the average
equivalent variation bound. As an example where it is nonlinear, they
discuss the causal mediation analysis of \citet{Imai.et.al.2010}.

A key feature of DML is the introduction of the Riesz representer
$\alpha_{0}\left(X\right).$ The Riesz representer is a function with
$\mathbb{E}\left[\alpha_{0}\left(X\right)^{2}\right]<\infty$ and
\begin{equation}
\mathbb{E}\left[m\left(W,\gamma\right)\right]=\mathbb{E}\left[\alpha_{0}\left(X\right)\boldsymbol{\gamma}\left(X\right)\right]\;\mathrm{for}\;\mathrm{all}\;\boldsymbol{\gamma}\;\mathrm{s}.\mathrm{t}.\;\mathbb{E}\left[\boldsymbol{\gamma}\left(X\right)^{2}\right]<\infty.\label{eq:Rr}
\end{equation}
As noted in \citet{Chernozhukov.et.al.2022a}, the Riesz representation
theorem states that the existence of such an $\alpha_{0}\left(X\right)$
is equivalent to $\mathbb{E}\left[m\left(W,\boldsymbol{\gamma}\right)\right]$
being a mean square continuous functional of $\boldsymbol{\gamma}.$
In other words, $\mathbb{E}\left[m\left(W,\boldsymbol{\gamma}\right)\right]\leq C\left\Vert \boldsymbol{\gamma}\right\Vert $
for all $\boldsymbol{\gamma}$ where $\left\Vert \boldsymbol{\gamma}\right\Vert =\sqrt{\mathbb{E}\left[\boldsymbol{\gamma}\left(X\right)^{2}\right]}$
and $C>0.$ In addition, the existence of $\alpha_{0}\left(X\right)$
implies that $\theta_{0}$ has a finite semiparametric variance bound
(\citet{Newey1994}, \citet{Hirshberg.and.Wager.2021}, and \citet{Chernozhukov.et.al.2022b}).

By equation \eqref{eq:Rr} and the law of iterated expectations, the
parameter of interest can be expressed in three ways:
\[
\theta_{0}=\mathbb{E}\left[m\left(W,\boldsymbol{\gamma}_{0}\right)\right]=\mathbb{E}\left[\alpha_{0}\left(X\right)\boldsymbol{\gamma}_{0}\left(X\right)\right]=\mathbb{E}\left[\alpha_{0}\left(X\right)Y\right].
\]
It is well-known that estimating $\theta_{0}$ by plugging an estimator
$\hat{\boldsymbol{\gamma}}$ of $\boldsymbol{\gamma}_{0}$ into $m\left(W,\boldsymbol{\gamma}\right)$
and using the sample analogue can result in large biases when $\hat{\boldsymbol{\gamma}}$
is a high-dimension estimator entailing regularization and/or model
selection (\citet{Chernozhukov.et.al.2017}, \citet{Chernozhukov.et.al.2018},
and \citet{Chernozhukov.et.al.2022c}). In order to deal with this
issue, DML uses an orthogonal moment function for $\theta_{0}.$ As
in \citet{Chernozhukov.et.al.2022c}, define $\boldsymbol{\gamma}\left(F\right)$
as the probability limit (plim) of $\hat{\boldsymbol{\gamma}}$ when
an observation $W$ has the cumulative distribution function (cdf)
$F.$ Many high-dimensional estimators, including Lasso, are constructed
from a sequence of regressors $X=\left(X_{1},X_{2},\cdots\right)$
and have the following form:
\[
\hat{\boldsymbol{\gamma}}\left(x\right)=\sum_{j=1}^{\infty}\hat{\beta}_{j}x_{j},\;\hat{\beta}_{j^{'}}\neq0\quad\mathrm{for}\;\mathrm{a}\;\mathrm{finite}\;\mathrm{number}\;\mathrm{of}\;j^{'},
\]
where $x=\left(x_{1},x_{2},\cdots\right)$ is a possible realization
of $X.$ As \citet{Chernozhukov.et.al.2022a} points out, if $\hat{\boldsymbol{\gamma}}$
is a linear combination of $X,$ $\boldsymbol{\gamma}\left(F\right)$
will also be a linear combination of $X,$ or at least $\boldsymbol{\gamma}\left(F\right)$
can be approximated by such a linear combination. In addition, if
the estimators are based on the least squares prediction of $Y,$
the following holds:
\begin{equation}
\boldsymbol{\gamma}\left(F\right)=\underset{\boldsymbol{\gamma}\in\Gamma}{\arg\min}\mathbb{E}_{F}\left[\left\{ Y-\boldsymbol{\gamma}\left(X\right)\right\} ^{2}\right]\label{eq:gamma}
\end{equation}
With properly defined $\Gamma,$ $\boldsymbol{\gamma}\left(F\right)$
becomes equivalent to $\mathbb{E}_{F}\left[Y\mid X\right].$ For example,
in Lasso, as long as $X=\left(X_{1},X_{2},\cdots\right)$ can approximate
any function of a fixed set of regressors, this is the case when $\Gamma$
is the set of all (measurable) functions of $X$ with finite second
moment (\citet{Chernozhukov.et.al.2022a}).

The orthogonal moment function from \citet{Chernozhukov.et.al.2022c}
is constructed by adding the nonparametric influence function of $\mathbb{E}\left[m\left(W,\boldsymbol{\gamma}\left(F\right)\right)\right]$
to the identifying moment function $m\left(w,\boldsymbol{\gamma}\right)-\theta.$
\citet{Newey1994} shows that the nonparametric influence function
of $\mathbb{E}\left[m\left(W,\boldsymbol{\gamma}\left(F\right)\right)\right]$
is
\[
\overline{\alpha}\left(X\right)\left[Y-\overline{\boldsymbol{\gamma}}\left(X\right)\right],
\]
where $\overline{\boldsymbol{\gamma}}\left(X\right)$ is the solution
to the equation \eqref{eq:gamma} for $F=F_{0}$ which is the (true)
cdf of $W,$ and $\overline{\alpha}\in\Gamma$ satisfies $\mathbb{E}\left[m\left(W,\boldsymbol{\gamma}\right)\right]=\mathbb{E}\left[\overline{\alpha}\left(X\right)\boldsymbol{\gamma}\left(X\right)\right]$
for all $\boldsymbol{\gamma}\in\Gamma.$ \citet{Chernozhukov.et.al.2022b}
shows that

\[
\overline{\alpha}=\underset{\alpha\in\Gamma}{\arg\min}\mathbb{E}\left[\left\{ \alpha_{0}\left(X\right)-\alpha\left(X\right)\right\} ^{2}\right].
\]
$\overline{\alpha}$ can be interpreted as the Riesz representer of
the linear functional $\mathbb{E}\left[m\left(W,\boldsymbol{\gamma}\right)\right]$
with domain $\Gamma.$ The orthogonal moment function is
\[
\psi\left(w,\theta,\boldsymbol{\gamma},\alpha\right)=m\left(w,\boldsymbol{\gamma}\right)-\theta+\alpha\left(x\right)\left[y-\boldsymbol{\gamma}\left(x\right)\right]
\]
From \citet{Chernozhukov.et.al.2022b}, for any $\boldsymbol{\gamma},\alpha\in\Gamma,$
\[
\mathbb{E}\left[\psi\left(W,\theta,\boldsymbol{\gamma},\alpha\right)-\psi\left(W,\theta,\overline{\boldsymbol{\gamma}},\overline{\alpha}\right)\right]=-\mathbb{E}\left[\left\{ \alpha\left(X\right)-\overline{\alpha}\left(X\right)\right\} \left\{ \boldsymbol{\gamma}\left(X\right)-\overline{\boldsymbol{\gamma}}\left(X\right)\right\} \right]
\]
and
\[
\mathbb{E}\left[\psi\left(W,\theta_{0},\overline{\boldsymbol{\gamma}},\overline{\alpha}\right)\right]=-\mathbb{E}\left[\left\{ \overline{\alpha}\left(X\right)-\alpha_{0}\left(X\right)\right\} \left\{ \overline{\boldsymbol{\gamma}}\left(X\right)-\boldsymbol{\gamma}_{0}\left(X\right)\right\} \right].
\]
Thus, $\mathbb{E}\left[\psi\left(W,\theta_{0},\overline{\boldsymbol{\gamma}},\overline{\alpha}\right)\right]=0$
if $\overline{\boldsymbol{\gamma}}=\boldsymbol{\gamma}_{0}$ or $\overline{\alpha}=\alpha_{0}.$
In other words, the orthogonal moment condition identifies $\theta_{0}$
when $\overline{\boldsymbol{\gamma}}\left(X\right)=\mathbb{E}_{F_{0}}\left[Y\mid X\right]$
or $\alpha_{0}\left(X\right)\in\Gamma.$

\citet{Chernozhukov.et.al.2022a} studies the case where $m\left(W,\boldsymbol{\gamma}\right)$
is nonlinear in $\boldsymbol{\gamma},$ and $\boldsymbol{\gamma}=\left(\gamma_{1}\left(X_{1}\right),\cdots\gamma_{K}\left(X_{K}\right)\right)^{'}$
with each regression $\gamma_{k}\left(X_{k}\right)$ using a specific
regressors $X_{k}.$ $m\left(W,\boldsymbol{\gamma}\right)$ is linearized
using Gateaux derivatives, and the Riesz representer is constructed
for each regression $\gamma_{k}\left(X_{k}\right).$ \citet{Chernozhukov.et.al.2022a}
shows the asymptotic normality of the DML estimator for both linear
and nonlinear cases.

DML involves cross-fitting where orthogonal moment functions are averaged
over observations different from those used to estimate $\overline{\gamma}$
and $\overline{\alpha}.$ It is known that cross-fitting removes a
source of bias and eliminates the need for Donsker conditions. Given
that many machine learning estimators do not satisfy Donsker conditions,
cross-fitting allows researchers to utilize these estimators (\citet{Chernozhukov.et.al.2018}).

\section{Proofs of Results}

\subsection{Proposition \ref{prop:1}}
\label{app1}

\begin{proof}
The proof of Proposition \ref{prop:1} mostly follows that of Theorem 9 of \citet{Chernozhukov.et.al.2022a}.
Theorem 9 derives the asymptotic distributions of the nonlinear DML
estimator under the Assumptions 1, 4, 5, 10, and 12-14. These seven
assumptions are used to verify Assumptions 1-3 in \citet{Chernozhukov.et.al.2022c}.
If Assumptions 1-3 were all satisfied, the following would hold:
\[
\sqrt{n}\left(\hat{\theta}_{\mathrm{sig}}-\overline{\theta}_{\mathrm{sig}}\right)=\frac{1}{\sqrt{n}}\sum_{i=1}^{n}\psi_{\mathrm{sig}}\left(W_{i},\overline{\boldsymbol{\gamma}},\overline{\alpha},\overline{\theta}_{\mathrm{sig}}\right)+o_{p}\left(1\right)
\]
This is not the case in our setting because Assumption 13 of \citet{Chernozhukov.et.al.2022a}
does not hold. In addition, condition (2) of their Assumption 12 is replaced by our Assumption 5. The remaining relevant conditions are verified below. We then show how the violation of Assumption 13(3) leads to a different conclusion.

\textbf{Assumption 1} For each $k=1,2,$ there exists $b_{k}\left(x_{k}\right)=\left(b_{k1}\left(x_{k}\right),\cdots,b_{kp}\left(x_{k}\right)\right)^{'}$
such that (1) $b_{kj}\in\Gamma_{k}$ for all $j=1,2,\cdots,p,$ and
(2) for any $\alpha_{k}\in\Gamma_{k}$ and $\epsilon_{k}>0,$ there
are $p$ and $\rho_{k}\in\mathbb{R}^{p}$ such that $\mathbb{E}\left[\alpha_{k}\left(X_{k}\right)-b_{k}\left(X_{k}\right)^{'}\rho_{k}\right]<\epsilon_{k}$
where $\Gamma_{k}$ is the set of each regression $\gamma_{k}\left(X_{k}\right).$

Assumption 1 implies that a linear combination of $b_k(x_k)$
approximates any element of $\Gamma_k$, while each element of
$b_k(x_k)$ belongs to $\Gamma_k$. In high-dimensional regression
applications, the raw covariates
$b_k(x_k)=(x_{k1},\ldots,x_{kp})'$
generally provide a sufficient basic dictionary, although additional
transformations may be included when needed.

\textbf{Assumption 4} For each $k=1,2,$ there exists $C_{k}>0$ such
that, with probability 1, $\sup_{j}\left|b_{kj}\left(X_{k}\right)\right|\leq C_{k}.$

Assumption 4 requires the elements of the dictionary
\(b_k(X_k)\) to be uniformly bounded. The raw-coordinate choice
satisfies this condition when the covariates have bounded support.
For unbounded covariates, including the Gaussian covariates used
in the simulations, analogous concentration results may instead
be obtained under appropriate tail conditions.

\textbf{Assumption 5} For each $k=1,2,$ $\epsilon_{n}=n^{-d_{\boldsymbol{\gamma}}},$
$r_{k}=o\left(n^{c}\epsilon_{n}\right)$ for all $c>0,$ and there
exists $C>0$ such that $p\leq Cn^{C}.$

Assumption 5 restricts the growth rate of $p$ to be slower than some
power of $n,$ and we accept it as a regularity condition.

\textbf{Assumption 10} $\mathbb{E}\left[m_{\mathrm{sig}}\left(W,\boldsymbol{\gamma}_{0}\right)\right]<\infty$
and $\int\left[m_{\mathrm{sig}}\left(w,\hat{\boldsymbol{\gamma}}\right)-m_{\mathrm{sig}}\left(w,\overline{\boldsymbol{\gamma}}\right)\right]F_{W}\left(dw\right)\overset{p}{\rightarrow}0.$

As in \citet{Chernozhukov.et.al.2022a}, Assumption 10 is implied
by the existence of $C>0$ with $\left|\mathbb{E}\left[m_{\mathrm{sig}}\left(W,\boldsymbol{\gamma}\right)^{2}\right]\right|\leq C\left\Vert \boldsymbol{\gamma}\right\Vert ^{2}$
for all $\boldsymbol{\gamma}.$ The inequality holds as follows:
\begin{eqnarray*}
\left|\mathbb{E}\left[m_{\mathrm{sig}}\left(W,\boldsymbol{\gamma}\right)^{2}\right]\right| & = & \left|\mathbb{E}\left[\left\{ \frac{\gamma_{1}\left(X\right)-\gamma_{2}\left(X\right)}{1+\exp\left(-s_{n}\left\{ \gamma_{1}\left(X\right)-\gamma_{2}\left(X\right)\right\} \right)}\right\} ^{2}\right]\right|\\
 & \leq & \left|\mathbb{E}\left[\left\{ \gamma_{1}\left(X\right)-\gamma_{2}\left(X\right)\right\} ^{2}\right]\right|\\
 & = & \left\Vert \gamma_{1}\left(X\right)-\gamma_{2}\left(X\right)\right\Vert ^{2}\\
 & \leq & \left(\left\Vert \gamma_{1}\left(X\right)\right\Vert +\left\Vert \gamma_{2}\left(X\right)\right\Vert \right)^{2}\\
 & \leq & C\left\Vert \boldsymbol{\gamma}\right\Vert ^{2}
\end{eqnarray*}
where the first inequality holds as the denominator is larger than
1, the second equality holds by the definition of the $L_{2}$-norm,
and the second inequality holds by the triangle inequality.

\textbf{Assumption 12} For $\tilde{\boldsymbol{\gamma}}=\left(\tilde{\gamma}_{1},\tilde{\gamma}_{2}\right)^{'}\in\prod_{k=1}^{2}\Gamma_{k}$
and $\gamma_{k}\in\Gamma_{k},$ define
\[
D_{k}\left(W,\gamma_{k},\tilde{\boldsymbol{\gamma}}\right)\equiv\frac{\partial m_{\mathrm{sig}}\left(W,\tilde{\boldsymbol{\gamma}}+e_{k}\eta\gamma_{k}\right)}{\partial\eta}\mid_{\eta=0}
\]
as the Gateaux derivative of $m_{\mathrm{sig}}\left(W,\boldsymbol{\gamma}\right)$
with respect to $\gamma_{k}$ where $e_{k}$ is the $k$th column
of the $2\times2$ identity matrix. Then, there are $C,$ $\epsilon>0,$$a_{kj}\left(w\right),$
and $A_{k}\left(w,\boldsymbol{\gamma}\right)$ such that, for all
$\boldsymbol{\gamma}$ with $\left\Vert \boldsymbol{\gamma}-\overline{\boldsymbol{\gamma}}\right\Vert \leq\epsilon,$
$D_{k}\left(W,b_{kj},\boldsymbol{\gamma}\right)$ exists and for $k=1,2,$
\begin{align*}
(1) & \;D_{k}\left(W,b_{kj},\boldsymbol{\gamma}\right)=a_{kj}\left(W\right)A_{k}\left(W,\boldsymbol{\gamma}\right)\\
(2) & \;\max_{j\leq p}\left|\mathbb{E}\left[a_{kj}\left(W\right)\left\{ A_{k}\left(W,\boldsymbol{\gamma}\right)-A_{k}\left(W,\overline{\boldsymbol{\gamma}}\right)\right\} \right]\right|\leq C\left\Vert \boldsymbol{\gamma}-\overline{\boldsymbol{\gamma}}\right\Vert \\
(3) & \;\max_{j\leq p}\left|a_{kj}\left(W\right)\right|\leq C\\
(4) & \;\mathbb{E}\left[A_{k}\left(W,\boldsymbol{\gamma}\right)^{2}\right]\leq C
\end{align*}

Assumption 12 imposes restrictions on the derivatives. (1) is satisfied
because $D_{1}\left(W,b_{1j},\boldsymbol{\gamma}\right)=a_{1j}\left(W\right)A_{1}\left(W,\boldsymbol{\gamma}\right)$
and $D_{2}\left(W,b_{2j},\boldsymbol{\gamma}\right)=a_{2j}\left(W\right)A_{2}\left(W,\boldsymbol{\gamma}\right)$
where
\begin{eqnarray*}
a_{1j}\left(W\right) & = & b_{1j}\\
a_{2j}\left(W\right) & = & b_{2j}\\
A_{1}\left(W,\boldsymbol{\gamma}\right) & = & \frac{1+\left\{ 1+s\left(\gamma_{1}-\gamma_{2}\right)\right\} e^{-s\left(\gamma_{1}-\gamma_{2}\right)}}{\left[1+e^{-s\left(\gamma_{1}-\gamma_{2}\right)}\right]^{2}}\\
A_{2}\left(W,\boldsymbol{\gamma}\right) & = & -A_{1}\left(W,\boldsymbol{\gamma}\right).
\end{eqnarray*}
(3) is satisfied as Assumption 4 of \citet{Chernozhukov.et.al.2022a}.
(4) is satisfied because $A_{k}\left(W,\boldsymbol{\gamma}\right)$ is uniformly bounded. Moreover, (2) is replaced by Assumption \ref{assu:5}, which directly imposes the required estimated-Riesz consistency and product-rate conditions.

\textbf{Assumption 14} There is $\frac{1}{4}<d_{\boldsymbol{\gamma}}<\frac{1}{2}$
such that $\left\Vert \hat{\gamma}_{k}-\overline{\gamma}_{k}\right\Vert =O_{p}\left(n^{-d_{\boldsymbol{\gamma}}}\right)$
for $k=1,2.$ Also, for each $\overline{\alpha}_{k}$ and $b_{k}\left(x_{k}\right),$
Assumptions 2 and 3 are satisfied with $\frac{d_{\boldsymbol{\gamma}}\left(1+4\xi\right)}{1+2\xi}>\frac{1}{2}$

We accept $\frac{1}{4}<d_{\boldsymbol{\gamma}}<\frac{1}{2}$ with
$\left\Vert \hat{\gamma}_{k}-\overline{\gamma}_{k}\right\Vert =O_{p}\left(n^{-d_{\boldsymbol{\gamma}}}\right)$
as a regularity condition. Assumptions 2 and 3 of \citet{Chernozhukov.et.al.2022a}
are verified as follows.

\textbf{Assumption 2} For each $k=1,2,$ there exists $C>0,$ $\xi>0$
such that for each positive integer $q\leq C\epsilon_{n}^{-\frac{2}{2\xi+1}},$
there is $\overline{\rho}_{k}$ with $q$ nonzero elements such that
\[
\left\Vert \overline{\alpha}_{k}-b_{k}^{'}\overline{\rho}_{k}\right\Vert \leq Cq^{-\xi}.
\]

We impose Assumption 2 as a high-level approximate-sparsity condition on the Riesz representers associated with the smoothing sequence \(s_n\). The corresponding consistency and product-rate requirements for the estimated Riesz representers are imposed directly in Assumption \ref{assu:5}.

\textbf{Assumption 3} For a matrix $A,$ define the norm $\left\Vert A\right\Vert _{1}=\sum_{i,j}\left|a_{ij}\right|.$
For a $p\times1$ vector $\rho,$ let $\rho_{J}$ be a $J\times1$
subvector of $\rho,$ and $\rho_{J^{c}}$ be the vector consisting
of all components of $\rho$ that are not in $\rho_{J}.$ Then, for
each $k=1,2,$ $G_{k}=\mathbb{E}\left[b_{k}\left(X_{k}\right)b_{k}\left(X_{k}\right)^{'}\right]$
has the largest eigenvalue bounded uniformly in $n$ and there are
$C,$ $c>0$ such that, for all $q\approx C\epsilon_{n}^{-2}$ with
probability approaching 1,
\[
\min_{J\leq q}\min_{\left\Vert \rho_{J^{c}}\right\Vert _{1}\leq3\left\Vert \rho_{J}\right\Vert _{1}}\frac{\rho^{'}\hat{G}_{k}\rho}{\rho_{J}^{'}\rho_{J}}\geq c.
\]

As in \citet{Chernozhukov.et.al.2022a}, Assumption 3 is a sparse
eigenvalue condition that is assumed in general in Lasso literature,
and we accept it as a regularity condition.

Unlike the above six assumptions, Assumption 13 is violated. In particular,
Assumption 13-(3) is violated due to the smoothing parameter $s_{n}.$

\textbf{Assumption 13} (1) For $k=1,2,$ there is $\overline{\alpha}_{k}\in\Gamma_{k}$
such that for all $\gamma_{k}\in\Gamma_{k},$ $\mathbb{E}\left[D_{k}\left(W,\gamma_{k},\overline{\boldsymbol{\gamma}}\right)\right]=\mathbb{E}\left[\overline{\alpha}_{k}\left(X_{k}\right)\gamma_{k}\left(X_{k}\right)\right];$
(2) $\overline{\alpha}_{k}\left(X_{k}\right)$ and $\mathbb{E}\left[\left\{ Y_{k}-\overline{\gamma}_{k}\left(X_{k}\right)\right\} ^{2}\mid X_{k}\right]$
are bounded; (3) there are $\epsilon,$ $C>0$ such that for all $\boldsymbol{\gamma}\in\prod_{k=1}^{2}\Gamma_{k}$
with $\left\Vert \boldsymbol{\gamma}-\overline{\boldsymbol{\gamma}}\right\Vert <\epsilon,$
\[
\left|\mathbb{E}\left[m_{\mathrm{sig}}\left(W,\boldsymbol{\gamma}\right)-m_{\mathrm{sig}}\left(W,\overline{\boldsymbol{\gamma}}\right)-\sum_{k=1}^{K}D_{k}\left(W,\gamma_{k}-\overline{\gamma}_{k},\overline{\boldsymbol{\gamma}}\right)\right]\right|\leq C\left\Vert \boldsymbol{\gamma}-\overline{\boldsymbol{\gamma}}\right\Vert ^{2}
\]

Assumption 13 shows that each $\overline{\alpha}_{k}$ can be viewed
as the Riesz representer for a linearized functional $\mathbb{E}\left[D_{k}\left(W,\gamma_{k},\overline{\boldsymbol{\gamma}}\right)\right],$
and the linearization error with respect to Gateaux derivatives is
bounded by a constant. By the definition of the Gateaux derivative,
\[
D_{1}\left(W,\gamma_{1},\overline{\boldsymbol{\gamma}}\right)=\frac{1+\left\{ 1+s_{n}\left(\overline{\gamma}_{1}-\overline{\gamma}_{2}\right)\right\} \exp\left(-s_{n}\left(\overline{\gamma}_{1}-\overline{\gamma}_{2}\right)\right)}{\left[1+\exp\left(-s_{n}\left(\overline{\gamma}_{1}-\overline{\gamma}_{2}\right)\right)\right]^{2}}\gamma_{1}.
\]
We notice that the Riesz representer for the first regression is
\[
\overline{\alpha}_{1}\left(X_{1}\right)=\frac{1+\left\{ 1+s_{n}\left(\overline{\gamma}_{1}-\overline{\gamma}_{2}\right)\right\} \exp\left(-s_{n}\left(\overline{\gamma}_{1}-\overline{\gamma}_{2}\right)\right)}{\left[1+\exp\left(-s_{n}\left(\overline{\gamma}_{1}-\overline{\gamma}_{2}\right)\right)\right]^{2}}\frac{D}{\mathrm{Pr}\left(D=1\mid Z\right)}
\]
because
\begin{eqnarray*}
    \mathbb{E}\left[\overline{\alpha}_{1}\left(X_{1}\right)\gamma_{1}\right] & = & \mathbb{E}\left[\frac{1+\left\{ 1+s_{n}\left(\overline{\gamma}_{1}-\overline{\gamma}_{2}\right)\right\} \exp\left(-s_{n}\left(\overline{\gamma}_{1}-\overline{\gamma}_{2}\right)\right)}{\left[1+\exp\left(-s_{n}\left(\overline{\gamma}_{1}-\overline{\gamma}_{2}\right)\right)\right]^{2}}\frac{D}{\mathrm{Pr}\left(D=1\mid Z\right)}\gamma_{1}\right]\\
    & = & \mathbb{E}\left[\frac{1+\left\{ 1+s_{n}\left(\overline{\gamma}_{1}-\overline{\gamma}_{2}\right)\right\} \exp\left(-s_{n}\left(\overline{\gamma}_{1}-\overline{\gamma}_{2}\right)\right)}{\left[1+\exp\left(-s_{n}\left(\overline{\gamma}_{1}-\overline{\gamma}_{2}\right)\right)\right]^{2}}\frac{\mathbb{E}\left[D\mid Z\right]}{\mathrm{Pr}\left(D=1\mid Z\right)}\gamma_{1}\right]\\
    & = & \mathbb{E}\left[\frac{1+\left\{ 1+s_{n}\left(\overline{\gamma}_{1}-\overline{\gamma}_{2}\right)\right\} \exp\left(-s_{n}\left(\overline{\gamma}_{1}-\overline{\gamma}_{2}\right)\right)}{\left[1+\exp\left(-s_{n}\left(\overline{\gamma}_{1}-\overline{\gamma}_{2}\right)\right)\right]^{2}}\gamma_{1}\right]\\
    & = & \mathbb{E}\left[D_{1}\left(W,\gamma_{1},\overline{\boldsymbol{\gamma}}\right)\right]
\end{eqnarray*}
Similarly, the Riesz representer for the second regression is
\[
\overline{\alpha}_{2}\left(X_{2}\right)=-\frac{1+\left\{ 1+s_{n}\left(\overline{\gamma}_{1}-\overline{\gamma}_{2}\right)\right\} \exp\left(-s_{n}\left(\overline{\gamma}_{1}-\overline{\gamma}_{2}\right)\right)}{\left[1+\exp\left(-s_{n}\left(\overline{\gamma}_{1}-\overline{\gamma}_{2}\right)\right)\right]^{2}}\frac{1-D}{1-\mathrm{Pr}\left(D=1\mid Z\right)}.
\]
so that (1) is satisfied. Next, (2) is satisfied because $\overline{\alpha}_{1}$ and $\overline{\alpha}_{2}$
are bounded, and the boundedness of $\mathbb{E}\left[\left\{ Y_{k}-\overline{\gamma}_{k}\left(X_{k}\right)^{2}\right\} \mid X_{k}\right]$
is given as a regularity condition. On the other hand, (3) is violated.
Given $\boldsymbol{\gamma}=\left[\begin{array}{c}
\gamma_{1}\\
\gamma_{2}
\end{array}\right],$ $\overline{\boldsymbol{\gamma}}=\left[\begin{array}{c}
\overline{\gamma}_{1}\\
\overline{\gamma}_{2}
\end{array}\right],$ for $\delta\in\left(0,1\right),$ the Taylor expansion yields

\begin{eqnarray*}
 &  & \mathbb{E}\left[m_{\mathrm{sig}}\left(W,\boldsymbol{\gamma}\right)-m_{\mathrm{sig}}\left(W,\overline{\boldsymbol{\gamma}}\right)-\sum_{k=1}^{K}D_{k}\left(W,\gamma_{k}-\overline{\gamma}_{k},\overline{\boldsymbol{\gamma}}\right)\right]\\
 & = & \frac{1}{2}\mathbb{E}\left[\left(\gamma_{1}-\overline{\gamma}_{1}\right)^{2}\frac{\partial^{2}}{\partial{\gamma}_{1}^{2}}m_{\mathrm{sig}}\left(W,\overline{\boldsymbol{\gamma}}+\delta\left(\boldsymbol{\gamma}-\overline{\boldsymbol{\gamma}}\right)\right)\right]\\
 &  & +\frac{1}{2}\mathbb{E}\left[\left(\gamma_{1}-\overline{\gamma}_{1}\right)\left(\gamma_{2}-\overline{\gamma}_{2}\right)\frac{\partial^{2}}{\partial{\gamma}_{1}\partial{\gamma}_{2}}m_{\mathrm{sig}}\left(W,\overline{\boldsymbol{\gamma}}+\delta\left(\boldsymbol{\gamma}-\overline{\boldsymbol{\gamma}}\right)\right)\right]\\
 &  & +\frac{1}{2}\mathbb{E}\left[\left(\gamma_{2}-\overline{\gamma}_{2}\right)\left(\gamma_{1}-\overline{\gamma}_{1}\right)\frac{\partial^{2}}{\partial{\gamma}_{2}\partial{\gamma}_{1}}m_{\mathrm{sig}}\left(W,\overline{\boldsymbol{\gamma}}+\delta\left(\boldsymbol{\gamma}-\overline{\boldsymbol{\gamma}}\right)\right)\right]\\
 &  & +\frac{1}{2}\mathbb{E}\left[\left(\gamma_{2}-\overline{\gamma}_{2}\right)^{2}\frac{\partial^{2}}{\partial{\gamma}_{2}^{2}}m_{\mathrm{sig}}\left(W,\overline{\boldsymbol{\gamma}}+\delta\left(\boldsymbol{\gamma}-\overline{\boldsymbol{\gamma}}\right)\right)\right]\\
 & = & \frac{1}{2}\mathbb{E}\left[a_{s}\left(\gamma_{1}-\overline{\gamma}_{1}\right)^{2}-2a_{s}\left(\gamma_{1}-\overline{\gamma}_{1}\right)\left(\gamma_{2}-\overline{\gamma}_{2}\right)+a_{s}\left(\gamma_{2}-\overline{\gamma}_{2}\right)^{2}\right]\\
 & = & \frac{1}{2}\mathbb{E}\left[a_{s}\left\{ \left(\gamma_{1}-\overline{\gamma}_{1}\right)-\left(\gamma_{2}-\overline{\gamma}_{2}\right)\right\} ^{2}\right]
\end{eqnarray*}
where
\begin{eqnarray*}
a_{s} & = & \frac{\partial^{2}}{\partial{\gamma}_{1}^{2}}m_{\mathrm{sig}}\left(W,\overline{\boldsymbol{\gamma}}+\delta\left(\boldsymbol{\gamma}-\overline{\boldsymbol{\gamma}}\right)\right)\\
 & = & \frac{\partial^{2}}{\partial{\gamma}_{2}^{2}}m_{\mathrm{sig}}\left(W,\overline{\boldsymbol{\gamma}}+\delta\left(\boldsymbol{\gamma}-\overline{\boldsymbol{\gamma}}\right)\right)\\
 & = & -\frac{\partial^{2}}{\partial{\gamma}_{1}\partial{\gamma}_{2}}m_{\mathrm{sig}}\left(W,\overline{\boldsymbol{\gamma}}+\delta\left(\boldsymbol{\gamma}-\overline{\boldsymbol{\gamma}}\right)\right)
\end{eqnarray*}
Note that
\[
\begin{aligned}
a_s
&=
2s_n
\exp\!\left(
-s_n\left\{
(1-\delta)(\overline{\gamma}_1-\overline{\gamma}_2)
+\delta(\gamma_1-\gamma_2)
\right\}
\right)\\
&\quad\times
\left[
1+
\exp\!\left(
-s_n\left\{
(1-\delta)(\overline{\gamma}_1-\overline{\gamma}_2)
+\delta(\gamma_1-\gamma_2)
\right\}
\right)
\right]^{-2}\\
&\quad+
s_n^2
\left\{
(1-\delta)(\overline{\gamma}_1-\overline{\gamma}_2)
+\delta(\gamma_1-\gamma_2)
\right\}\\
&\quad\times
\exp\!\left(
-s_n\left\{
(1-\delta)(\overline{\gamma}_1-\overline{\gamma}_2)
+\delta(\gamma_1-\gamma_2)
\right\}
\right)\\
&\quad\times
\left[
1+
\exp\!\left(
-s_n\left\{
(1-\delta)(\overline{\gamma}_1-\overline{\gamma}_2)
+\delta(\gamma_1-\gamma_2)
\right\}
\right)
\right]^{-3}\\
&\quad\times
\Bigg[
2\exp\!\left(
-s_n\left\{
(1-\delta)(\overline{\gamma}_1-\overline{\gamma}_2)
+\delta(\gamma_1-\gamma_2)
\right\}
\right)\\
&\hspace{5em}
-\left\{
1+
\exp\!\left(
-s_n\left\{
(1-\delta)(\overline{\gamma}_1-\overline{\gamma}_2)
+\delta(\gamma_1-\gamma_2)
\right\}
\right)
\right\}
\Bigg].
\end{aligned}
\]
The final bracket simplifies to
\[
\exp\!\left(
-s_n\left\{
(1-\delta)(\overline{\gamma}_1-\overline{\gamma}_2)
+\delta(\gamma_1-\gamma_2)
\right\}
\right)-1.
\]
For notational convenience, let $z\equiv\left(1-\delta\right)\left(\overline{\gamma}_{1}-\overline{\gamma}_{2}\right)+\delta\left(\gamma_{1}-\gamma_{2}\right).$ Then,
\begin{eqnarray*}
a_{s} & = & \frac{2\exp\left(-s_{n}z\right)}{\left[1+\exp\left(-s_{n}z\right)\right]^{2}}s_{n}\\
 &  & +\frac{\left(-s_{n}z\right)\exp\left(-s_{n}z\right)\left[1-\exp\left(-s_{n}z\right)\right]}{\left[1+\exp\left(-s_{n}z\right)\right]^{3}}s_{n}\\
 & = & s_{n}\left[\frac{2\exp\left(-s_{n}z\right)}{\left[1+\exp\left(-s_{n}z\right)\right]^{2}}+\frac{\left(-s_{n}z\right)\exp\left(-s_{n}z\right)\left[1-\exp\left(-s_{n}z\right)\right]}{\left[1+\exp\left(-s_{n}z\right)\right]^{3}}\right]
\end{eqnarray*}
Since
\[
\left|\frac{2\exp\left(-s_{n}z\right)}{\left[1+\exp\left(-s_{n}z\right)\right]^{2}}+\frac{\left(-s_{n}z\right)\exp\left(-s_{n}z\right)\left[1-\exp\left(-s_{n}z\right)\right]}{\left[1+\exp\left(-s_{n}z\right)\right]^{3}}\right|\leq\frac{1}{2}
\]
we obtain
\begin{eqnarray*}
\left|a_{s}\right| & = & s_{n}\left|\frac{2\exp\left(-s_{n}z\right)}{\left[1+\exp\left(-s_{n}z\right)\right]^{2}}+\frac{\left(-s_{n}z\right)\exp\left(-s_{n}z\right)\left[1-\exp\left(-s_{n}z\right)\right]}{\left[1+\exp\left(-s_{n}z\right)\right]^{3}}\right|\\
 & \leq & \frac{1}{2}s_{n}
\end{eqnarray*}
Therefore,
\begin{eqnarray*}
 &  & \left|\mathbb{E}\left[m_{\mathrm{sig}}\left(W,\boldsymbol{\gamma}\right)-m_{\mathrm{sig}}\left(W,\overline{\boldsymbol{\gamma}}\right)-\sum_{k=1}^{K}D_{k}\left(W,\gamma_{k}-\overline{\gamma}_{k},\overline{\boldsymbol{\gamma}}\right)\right]\right|\\
 & = & \frac{1}{2}\left|\mathbb{E}\left[a_{s}\left\{ \left(\gamma_{1}-\overline{\gamma}_{1}\right)-\left(\gamma_{2}-\overline{\gamma}_{2}\right)\right\} ^{2}\right]\right|\\
 & \leq & \frac{1}{2}\left|a_{s}\right|\mathbb{E}\left[\left\{ \left(\gamma_{1}-\overline{\gamma}_{1}\right)-\left(\gamma_{2}-\overline{\gamma}_{2}\right)\right\} ^{2}\right]\\
 & \leq & \frac{1}{4}s_{n}\left\Vert \left(\gamma_{1}-\overline{\gamma}_{1}\right)-\left(\gamma_{2}-\overline{\gamma}_{2}\right)\right\Vert ^{2}\\
 & \leq & \frac{1}{2}s_{n}\left(\left\Vert \gamma_{1}-\overline{\gamma}_{1}\right\Vert ^{2}+\left\Vert \gamma_{2}-\overline{\gamma}_{2}\right\Vert ^{2}\right)\\
 & = & \frac{1}{2}s_{n}\left\Vert \boldsymbol{\gamma}-\overline{\boldsymbol{\gamma}}\right\Vert ^{2}\\
 & = & C\left(s_{n}\right)\left\Vert \boldsymbol{\gamma}-\overline{\boldsymbol{\gamma}}\right\Vert ^{2}
\end{eqnarray*}
The bound of the remainder term thus involves a quantity $C\left(s_{n}\right)$
which depends on the smoothing parameter $s_{n}.$ The rest of the
proof involves generalizing the proof of \citet{Chernozhukov.et.al.2022c}
to cases where Assumption 13-(3) is violated. Define
\begin{eqnarray*}
\phi_{k}\left(w,\gamma_{k},\alpha_{k}\right) & \equiv & \alpha_{k}\left(x_{k}\right)\left[y_{k}-\gamma_{k}\left(x_{k}\right)\right]\\
g\left(w,\boldsymbol{\gamma},\theta\right) & \equiv & m_{\mathrm{sig}}\left(w,\boldsymbol{\gamma}\right)-\theta\\
\phi\left(w,\boldsymbol{\gamma},\boldsymbol{\alpha}\right) & \equiv & \sum_{k=1}^{2}\phi_{k}\left(w,\gamma_{k},\alpha_{k}\right)
\end{eqnarray*}
Also, define
\begin{eqnarray*}
\psi_{\mathrm{sig}}\left(w,\boldsymbol{\gamma},\boldsymbol{\alpha},\theta\right) & \equiv & m_{\mathrm{sig}}\left(w,\boldsymbol{\gamma}\right)-\theta+\phi\left(w,\boldsymbol{\gamma},\boldsymbol{\alpha}\right)\\
 & = & g\left(w,\boldsymbol{\gamma},\theta\right)+\phi\left(w,\boldsymbol{\gamma},\boldsymbol{\alpha}\right)
\end{eqnarray*}
Subtracting $\overline{\theta}_{\mathrm{sig}}$ from both sides of
equation \eqref{eq:estimator} gives
\begin{eqnarray*}
\hat{\theta}_{\mathrm{sig}}-\overline{\theta}_{\mathrm{sig}} & = & \frac{1}{n}\sum_{l=1}^{L}\sum_{i\in I_{l}}\left\{ m_{\mathrm{sig}}\left(W_{i},\hat{\boldsymbol{\gamma}}_{l}\right)-\overline{\theta}_{\mathrm{sig}}+\sum_{k=1}^{2}\hat{\alpha}_{kl}\left(X_{ki}\right)\left[Y_{ki}-\hat{\gamma}_{kl}\left(X_{ki}\right)\right]\right\} \\
 & = & \frac{1}{n}\sum_{l=1}^{L}\sum_{i\in I_{l}}\left\{ g\left(W_{i},\hat{\boldsymbol{\gamma}}_{l},\overline{\theta}_{\mathrm{sig}}\right)+\phi\left(W_{i},\hat{\boldsymbol{\gamma}}_{l},\hat{\boldsymbol{\alpha}}_{l}\right)\right\} \\
 & = & \frac{1}{n}\sum_{l=1}^{L}\sum_{i\in I_{l}}\left\{ \underset{\equiv\hat{R}_{1li}\left(W_{i}\right)}{\underbrace{g\left(W_{i},\hat{\boldsymbol{\gamma}}_{l},\overline{\theta}_{\mathrm{sig}}\right)-g\left(w,\overline{\boldsymbol{\gamma}},\overline{\theta}_{\mathrm{sig}}\right)}}+g\left(w,\overline{\boldsymbol{\gamma}},\overline{\theta}_{\mathrm{sig}}\right)\right\} \\
 &  & +\frac{1}{n}\sum_{l=1}^{L}\sum_{i\in I_{l}}\left\{ \phi\left(W_{i},\hat{\boldsymbol{\gamma}}_{l},\hat{\boldsymbol{\alpha}}_{l}\right)+\phi\left(W_{i},\overline{\boldsymbol{\gamma}},\overline{\boldsymbol{\alpha}}\right)-\phi\left(W_{i},\overline{\boldsymbol{\gamma}},\overline{\boldsymbol{\alpha}}\right)\right\} \\
 & = & \frac{1}{n}\sum_{l=1}^{L}\sum_{i\in I_{l}}\left\{ \hat{R}_{1li}+g\left(w,\overline{\boldsymbol{\gamma}},\overline{\theta}_{\mathrm{sig}}\right)\right\} \\
 &  & +\frac{1}{n}\sum_{l=1}^{L}\sum_{i\in I_{l}}\left\{ \underset{\equiv\hat{\Delta}_{l}\left(W_{i}\right)}{\underbrace{\phi\left(W_{i},\hat{\boldsymbol{\gamma}}_{l},\hat{\boldsymbol{\alpha}}_{l}\right)-\phi\left(W_{i},\overline{\boldsymbol{\gamma}},\hat{\boldsymbol{\alpha}}_{l}\right)-\phi\left(W_{i},\hat{\boldsymbol{\gamma}}_{l},\overline{\boldsymbol{\alpha}}\right)+\phi\left(W_{i},\overline{\boldsymbol{\gamma}},\overline{\boldsymbol{\alpha}}\right)}}\right\} \\
 &  & +\frac{1}{n}\sum_{l=1}^{L}\sum_{i\in I_{l}}\left\{ \phi\left(W_{i},\hat{\boldsymbol{\gamma}}_{l},\overline{\boldsymbol{\alpha}}\right)+\phi\left(W_{i},\overline{\boldsymbol{\gamma}},\hat{\boldsymbol{\alpha}}_{l}\right)-\phi\left(W_{i},\overline{\boldsymbol{\gamma}},\overline{\boldsymbol{\alpha}}\right)\right\} \\
 & = & \frac{1}{n}\sum_{l=1}^{L}\sum_{i\in I_{l}}\left\{ \hat{R}_{1li}+\hat{\Delta}_{l}\left(W_{i}\right)+g\left(w,\overline{\boldsymbol{\gamma}},\overline{\theta}_{\mathrm{sig}}\right)\right\} \\
 &  & +\frac{1}{n}\sum_{l=1}^{L}\sum_{i\in I_{l}}\left\{ \underset{\equiv\hat{R}_{2li}}{\underbrace{\phi\left(W_{i},\hat{\boldsymbol{\gamma}}_{l},\overline{\boldsymbol{\alpha}}\right)-\phi\left(W_{i},\overline{\boldsymbol{\gamma}},\overline{\boldsymbol{\alpha}}\right)}}+\underset{\equiv\hat{R}_{3li}}{\underbrace{\phi\left(W_{i},\overline{\boldsymbol{\gamma}},\hat{\boldsymbol{\alpha}}_{l}\right)-\phi\left(W_{i},\overline{\boldsymbol{\gamma}},\overline{\boldsymbol{\alpha}}\right)}}+\phi\left(W_{i},\overline{\boldsymbol{\gamma}},\overline{\boldsymbol{\alpha}}\right)\right\} \\
 & = & \frac{1}{n}\sum_{l=1}^{L}\sum_{i\in I_{l}}\left\{ \hat{R}_{1li}+\hat{R}_{2li}+\hat{R}_{3li}+\hat{\Delta}_{l}\left(W_{i}\right)+\psi_{\mathrm{sig}}\left(W_{i},\overline{\boldsymbol{\gamma}},\overline{\boldsymbol{\alpha}},\overline{\theta}_{\mathrm{sig}}\right)\right\} 
\end{eqnarray*}
Multiplying both sides by $\sqrt{n}$ gives
\begin{eqnarray*}
\sqrt{n}\left(\hat{\theta}_{\mathrm{sig}}-\overline{\theta}_{\mathrm{sig}}\right) & = & \frac{1}{\sqrt{n}}\sum_{l=1}^{L}\sum_{i\in I_{l}}\left\{ \hat{R}_{1li}+\hat{R}_{2li}+\hat{R}_{3li}\right\} \\
 &  & +\frac{1}{\sqrt{n}}\sum_{l=1}^{L}\sum_{i\in I_{l}}\hat{\Delta}_{l}\left(W_{i}\right)\\
 &  & +\frac{1}{\sqrt{n}}\sum_{i=1}^{n}\psi_{\mathrm{sig}}\left(W_{i},\overline{\boldsymbol{\gamma}},\overline{\boldsymbol{\alpha}},\overline{\theta}_{\mathrm{sig}}\right)
\end{eqnarray*}
If Assumptions 1, 2, and 3 of \citet{Chernozhukov.et.al.2022c} all
held, we would be able to show
\begin{eqnarray*}
\frac{1}{\sqrt{n}}\sum_{l=1}^{L}\sum_{i\in I_{l}}\left\{ \hat{R}_{1li}+\hat{R}_{2li}+\hat{R}_{3li}\right\}  & = & o_{p}\left(1\right)\\
\frac{1}{\sqrt{n}}\sum_{l=1}^{L}\sum_{i\in I_{l}}\hat{\Delta}_{l}\left(W_{i}\right) & = & o_{p}\left(1\right)
\end{eqnarray*}
so that
\begin{eqnarray*}
\sqrt{n}\left(\hat{\theta}_{\mathrm{sig}}-\overline{\theta}_{\mathrm{sig}}\right) & = & \frac{1}{\sqrt{n}}\sum_{i=1}^{n}\psi_{\mathrm{sig}}\left(W_{i},\overline{\boldsymbol{\gamma}},\overline{\boldsymbol{\alpha}},\overline{\theta}_{\mathrm{sig}}\right)+o_{p}\left(1\right)\\
 & \overset{d}{\rightarrow} & \mathcal{N}\left(0,V\right)
\end{eqnarray*}
where $V=\mathrm{Var}\left(\psi_{\mathrm{sig}}\left(W_{i},\overline{\boldsymbol{\gamma}},\overline{\boldsymbol{\alpha}},\overline{\theta}_{\mathrm{sig}}\right)\right).$
In our setting, the conclusion is different.

First, let us focus on $\frac{1}{\sqrt{n}}\sum_{l=1}^{L}\sum_{i\in I_{l}}\hat{\Delta}_{l}\left(W_{i}\right).$
Note that
\begin{eqnarray*}
\hat{\Delta}_{l}\left(W_{i}\right) & = & \phi\left(W_{i},\hat{\boldsymbol{\gamma}}_{l},\hat{\boldsymbol{\alpha}}_{l}\right)-\phi\left(W_{i},\overline{\boldsymbol{\gamma}},\hat{\boldsymbol{\alpha}}_{l}\right)-\phi\left(W_{i},\hat{\boldsymbol{\gamma}}_{l},\overline{\boldsymbol{\alpha}}\right)+\phi\left(W_{i},\overline{\boldsymbol{\gamma}},\overline{\boldsymbol{\alpha}}\right)\\
 & = & \sum_{k=1}^{2}\left(\phi_{k}\left(w,\hat{\gamma}_{kl},\hat{\alpha}_{kl}\right)-\phi_{k}\left(w,\overline{\gamma}_{k},\hat{\alpha}_{kl}\right)-\phi_{k}\left(w,\hat{\gamma}_{kl},\overline{\alpha}_{k}\right)+\phi_{k}\left(w,\overline{\gamma}_{k},\overline{\alpha}_{k}\right)\right)\\
 & = & \sum_{k=1}^{2}\left(-\hat{\alpha}_{kl}\hat{\gamma}_{kl}+\hat{\alpha}_{kl}\overline{\gamma}_{k}+\overline{\alpha}_{k}\hat{\gamma}_{kl}-\overline{\alpha}_{k}\overline{\gamma}_{k}\right)\\
 & = & -\sum_{k=1}^{2}\left[\left(\hat{\alpha}_{kl}-\overline{\alpha}_{k}\right)\left(\hat{\gamma}_{kl}-\overline{\gamma}_{k}\right)\right]
\end{eqnarray*}
and
\begin{eqnarray*}
\left|\frac{1}{\sqrt{n}}\sum_{i\in I_{l}}
\hat{\Delta}_{l}\left(W_{i}\right)\right|
& = &
\left|-\sum_{k=1}^{2}\frac{1}{\sqrt{n}}
\sum_{i\in I_{l}}
\left\{\hat{\alpha}_{kl}\left(X_{ki}\right)
-\overline{\alpha}_{k}\left(X_{ki}\right)\right\}
\left\{\hat{\gamma}_{kl}\left(X_{ki}\right)
-\overline{\gamma}_{k}\left(X_{ki}\right)\right\}\right|\\
& \leq &
\sum_{k=1}^{2}\sqrt{n}
\sqrt{\sum_{i\in I_{l}}
\frac{\left\{\hat{\alpha}_{kl}\left(X_{ki}\right)
-\overline{\alpha}_{k}\left(X_{ki}\right)\right\}^{2}}{n}}
\sqrt{\sum_{i\in I_{l}}
\frac{\left\{\hat{\gamma}_{kl}\left(X_{ki}\right)
-\overline{\gamma}_{k}\left(X_{ki}\right)\right\}^{2}}{n}}\\
& = &
O_{p}\left(
\sqrt{n}\sum_{k=1}^{2}
\left\Vert\hat{\alpha}_{kl}-\overline{\alpha}_{k}\right\Vert
\left\Vert\hat{\gamma}_{kl}-\overline{\gamma}_{k}\right\Vert
\right)\\
& = & o_{p}\left(1\right).
\end{eqnarray*}
where the results for $\left\Vert \hat{\alpha}_{kl}-\overline{\alpha}_{k}\right\Vert $
and $\left\Vert \hat{\gamma}_{kl}-\overline{\gamma}_{k}\right\Vert $
follow from Assumption \ref{assu:5}.

Second, to see the asymptotic behavior of $\frac{1}{\sqrt{n}}\sum_{i=1}^{n}\psi_{\mathrm{sig}}\left(W_{i},\overline{\boldsymbol{\gamma}},\overline{\boldsymbol{\alpha}},\overline{\theta}_{\mathrm{sig}}\right),$
note that
\begin{eqnarray*}
\psi_{\mathrm{sig}}\left(w_{i}\right) & \equiv & \psi_{\mathrm{sig}}\left(W_{i},\overline{\boldsymbol{\gamma}},\overline{\boldsymbol{\alpha}},\overline{\theta}_{\mathrm{sig}}\right)\\
 & = & m_{\mathrm{sig}}\left(W_{i},\overline{\boldsymbol{\gamma}}\right)-\overline{\theta}_{\mathrm{sig}}+\sum_{k=1}^{2}\overline{\alpha}_{k}\left(x_{ki}\right)\left[y_{ki}-\overline{\gamma}_{k}\left(x_{ki}\right)\right]\\
 & = & \frac{\overline{\tau}\left(Z_{i}\right)}{1+\exp\left(-s_{n}\overline{\tau}\left(Z_{i}\right)\right)}-\overline{\theta}_{\mathrm{sig}}+\sum_{k=1}^{2}\overline{\alpha}_{k}\left(x_{ki}\right)\left[y_{ki}-\overline{\gamma}_{k}\left(x_{ki}\right)\right].
\end{eqnarray*}
For notational convenience, define
\[
\Lambda_{n}\left(t\right)\equiv\frac{1}{1+\exp\left(-s_{n}t\right)}
\]
and
\[
m_{n}\left(t\right)\equiv t\Lambda_{n}\left(t\right).
\]
Then,
\[
m_{n}^{'}\left(t\right)=\Lambda_{n}\left(t\right)+s_{n}t\Lambda_{n}\left(t\right)\left(1-\Lambda_{n}\left(t\right)\right)\equiv q_{n}\left(t\right).
\]
We can rewrite the oracle orthogonal moment as
\[
\psi_{\mathrm{sig}}\left(W_{i}\right)=m_{n}\left(\overline{\tau}\left(Z_{i}\right)\right)-\overline{\theta}_{\mathrm{sig}}+\sum_{k=1}^{2}\overline{\alpha}_{k}\left(X_{ki}\right)\left[Y_{ki}-\overline{\gamma}_{k}\left(X_{ki}\right)\right]
\]
where
\begin{eqnarray*}
    \overline{\alpha}_{1}\left(X_{i}\right) & = & q_{n}\left(\overline{\tau}\left(Z_{i}\right)\right)\frac{D_{i}}{e\left(Z_{i}\right)}\\
    \overline{\alpha}_{2}\left(X_{i}\right) & = & -q_{n}\left(\overline{\tau}\left(Z_{i}\right)\right)\frac{1-D_{i}}{1-e\left(Z_{i}\right)}
\end{eqnarray*}
where $e\left(Z_{i}\right)=\mathrm{Pr}\left(D_{i}=1\mid Z_{i}\right)$.

We first characterize the limiting variance of $\psi_{\mathrm{sig}}\left(W_{i}\right)$. By the law of total variance, we have
\[
\mathrm{Var}\left(\psi_{\mathrm{sig}}\left(W_{i}\right)\right)=\mathrm{Var}\left(\mathbb{E}\left[\psi_{\mathrm{sig}}\left(W_{i}\right)\mid Z_{i}\right]\right)+\mathbb{E}\left[\mathrm{Var}\left(\psi_{\mathrm{sig}}\left(W_{i}\right)\mid Z_{i}\right)\right].
\]
Since
\[
\mathbb{E}\left[\frac{D_{i}}{e\left(Z_{i}\right)}\left\{ Y_{i}-\gamma_{1}\left(Z_{i}\right)\right\} -\frac{1-D_{i}}{1-e\left(Z_{i}\right)}\left\{ Y_{i}-\gamma_{2}\left(Z_{i}\right)\right\} \mid Z_{i}\right]=0,
\]
we have
\[
\mathbb{E}\left[\psi_{\mathrm{sig}}\left(W_{i}\right)\mid Z_{i}\right]=m_{n}\left(\overline{\tau}\left(Z_{i}\right)\right)-\overline{\theta}_{\mathrm{sig}}.
\]
Hence,
\[
\mathrm{Var}\left(\mathbb{E}\left[\psi_{\mathrm{sig}}\left(W_{i}\right)\mid Z_{i}\right]\right)=\mathrm{Var}\left(m_{n}\left(\overline{\tau}\left(Z_{i}\right)\right)\right).
\]
Next,
\[
\mathrm{Var}\left(\psi_{\mathrm{sig}}\left(W_{i}\right)\mid Z_{i}\right)=q_{n}\left(\overline{\tau}\left(Z_{i}\right)\right)^{2}\mathrm{Var}\left(\frac{D_{i}}{e\left(Z_{i}\right)}\left\{ Y_{i}-\gamma_{1}\left(Z_{i}\right)\right\} -\frac{1-D_{i}}{1-e\left(Z_{i}\right)}\left\{ Y_{i}-\gamma_{2}\left(Z_{i}\right)\right\} \mid Z_{i}\right).
\]
Because $D_{i}\left(1-D_{i}\right)=0$, the cross-product vanishes. Thus,
\[
\mathrm{Var}\left(\psi_{\mathrm{sig}}\left(W_{i}\right)\mid Z_{i}\right)=q_{n}\left(\overline{\tau}\left(Z_{i}\right)\right)^{2}\mathbb{E}\left[\frac{D_{i}}{e\left(Z_{i}\right)^{2}}\left\{ Y_{i}-\gamma_{1}\left(Z_{i}\right)\right\} ^{2}+\frac{1-D_{i}}{\left(1-e\left(Z_{i}\right)\right)^{2}}\left\{ Y_{i}-\gamma_{2}\left(Z_{i}\right)\right\} ^{2}\mid Z_{i}\right].
\]
Note that
\begin{eqnarray*}
    \mathbb{E}\left[\frac{D_{i}}{e\left(Z_{i}\right)^{2}}\left\{ Y_{i}-\gamma_{1}\left(Z_{i}\right)\right\} ^{2}\mid Z_{i}\right] & = & \frac{1}{e\left(Z_{i}\right)^{2}}\mathbb{E}\left[D_{i}\left\{ Y_{i}-\gamma_{1}\left(Z_{i}\right)\right\} ^{2}\mid Z_{i}\right]\\
    & = & \frac{1}{e\left(Z_{i}\right)^{2}}\mathrm{Pr}\left(D_{i}=1\mid Z_{i}\right)\mathbb{E}\left[\left\{ Y_{i}-\gamma_{1}\left(Z_{i}\right)\right\} ^{2}\mid D_{i}=1,Z_{i}\right]\\
    & = & \frac{\mathbb{E}\left[\left\{ Y_{i}-\gamma_{1}\left(Z_{i}\right)\right\} ^{2}\mid D_{i}=1,Z_{i}\right]}{e\left(Z_{i}\right)}\\
    & = & \frac{\sigma_{1}^{2}\left(Z_{i}\right)}{e\left(Z_{i}\right)}.
\end{eqnarray*}
Similarly,
\[
\mathbb{E}\left[\frac{1-D_{i}}{\left(1-e\left(Z_{i}\right)\right)^{2}}\left\{ Y_{i}-\gamma_{2}\left(Z_{i}\right)\right\} ^{2}\mid Z_{i}\right]=\frac{\sigma_{2}^{2}\left(Z_{i}\right)}{1-e\left(Z_{i}\right)}.
\]
Therefore,
\[
\mathbb{E}\left[\mathrm{Var}\left(\psi_{\mathrm{sig}}\left(W_{i}\right)\mid Z_{i}\right)\right]=\mathbb{E}\left[q_{n}\left(\tau\left(Z_{i}\right)\right)^{2}\left\{ \frac{\sigma_{1}^{2}\left(Z_{i}\right)}{e\left(Z_{i}\right)}+\frac{\sigma_{2}^{2}\left(Z_{i}\right)}{1-e\left(Z_{i}\right)}\right\} \right].
\]
Combining both terms,
\[
\mathrm{Var}\left(\psi_{\mathrm{sig}}\left(W_{i}\right)\right)=\mathrm{Var}\left(m_{n}\left(\overline{\tau}\left(Z_{i}\right)\right)\right)+\mathbb{E}\left[q_{n}\left(\overline{\tau}\left(Z_{i}\right)\right)^{2}\left\{ \frac{\sigma_{1}^{2}\left(Z_{i}\right)}{e\left(Z_{i}\right)}+\frac{\sigma_{2}^{2}\left(Z_{i}\right)}{1-e\left(Z_{i}\right)}\right\} \right].
\]
Since $s_{n}\rightarrow\infty$, We have
\[
m_{n}\left(\tau\right)\rightarrow\tau\mathds{1}\left\{ \tau>0\right\}.
\]
Moreover,
\[
q_{n}\left(\tau\right)=\Lambda_{n}\left(\tau\right)+s_{n}t\Lambda_{n}\left(\tau\right)\left(1-\Lambda_{n}\left(\tau\right)\right)\rightarrow q_{\infty}\left(\tau\right),
\]
where
\[
q_{\infty}\left(\tau\right)\equiv\mathds{1}\left\{ \tau>0\right\} +\frac{1}{2}\mathds{1}\left\{ \tau=0\right\}.
\]
Indeed, if $\tau>0$, then $\Lambda_{n}\left(\tau\right)\rightarrow1$ and $s_{n}t\Lambda_{n}\left(\tau\right)\left(1-\Lambda_{n}\left(\tau\right)\right)\rightarrow0$. If $\tau<0$, then $\Lambda_{n}\left(\tau\right)\rightarrow0$ and $s_{n}t\Lambda_{n}\left(\tau\right)\left(1-\Lambda_{n}\left(\tau\right)\right)\rightarrow0$. If $\tau=0$, then $\Lambda_{n}\left(0\right)=\frac{1}{2}$ and the second term is zero.

In addition, $m_{n}\left(\tau\right)$ is dominated by $\left|\tau\right|$, and $q_{n}\left(\tau\right)$ is uniformly bounded. Hence, under the finite moment and overlap conditions, the dominated convergence theorem gives
\[
\lim_{n\rightarrow\infty}\mathrm{Var}\left(\psi_{\mathrm{sig}}\left(W_{i}\right)\right)=V,
\]
where
\[
V=\mathrm{Var}\left(\overline{\tau}\left(Z_{i}\right)\mathds{1}\left\{ \overline{\tau}\left(Z_{i}\right)>0\right\} \right)+\mathbb{E}\left[q_{\infty}\left(\tau\right)^{2}\left\{ \frac{\sigma_{1}^{2}\left(Z_{i}\right)}{e\left(Z_{i}\right)}+\frac{\sigma_{2}^{2}\left(Z_{i}\right)}{1-e\left(Z_{i}\right)}\right\} \right].
\]
Equivalently,
\begin{eqnarray*}
    V & = & \mathrm{Var}\left(\overline{\tau}\left(Z_{i}\right)\mathds{1}\left\{ \overline{\tau}\left(Z_{i}\right)>0\right\} \right)+\mathbb{E}\left[\mathds{1}\left\{ \overline{\tau}\left(Z_{i}\right)>0\right\} \left\{ \frac{\sigma_{1}^{2}\left(Z_{i}\right)}{e\left(Z_{i}\right)}+\frac{\sigma_{2}^{2}\left(Z_{i}\right)}{1-e\left(Z_{i}\right)}\right\} \right]\\
    & & +\frac{1}{4}\mathbb{E}\left[\mathds{1}\left\{ \overline{\tau}\left(Z_{i}\right)=0\right\} \left\{ \frac{\sigma_{1}^{2}\left(Z_{i}\right)}{e\left(Z_{i}\right)}+\frac{\sigma_{2}^{2}\left(Z_{i}\right)}{1-e\left(Z_{i}\right)}\right\} \right].
\end{eqnarray*}
The remaining step is to verify the conditions of the Lyapunov central
limit theorem. Define 
\[
Q_{n}\equiv\sum_{i=1}^{n}\psi_{\mathrm{sig}}\left(W_{i},\overline{\boldsymbol{\gamma}},\overline{\boldsymbol{\alpha}},\overline{\theta}_{\mathrm{sig}}\right)
\]
to have
\begin{eqnarray*}
\mathbb{E}\left[Q_{n}\right] & = & \sum_{i=1}^{n}\mathbb{E}\left[\psi_{\mathrm{sig}}\left(W_{i},\overline{\boldsymbol{\gamma}},\overline{\boldsymbol{\alpha}},\overline{\theta}_{\mathrm{sig}}\right)\right]\\
 & = & 0\\
\mathrm{Var}\left(Q_{n}\right) & = & n\cdot\mathrm{Var}\left(\psi_{\mathrm{sig}}\left(W_{1},\overline{\boldsymbol{\gamma}},\overline{\boldsymbol{\alpha}},\overline{\theta}_{\mathrm{sig}}\right)\right).
\end{eqnarray*}
Construct
\[
\frac{Q_{n}-\mathbb{E}\left[Q_{n}\right]}{\sqrt{\mathrm{Var}\left(Q_{n}\right)}}=\sum_{i=1}^{n}L_{ni}
\]
where
\[
L_{ni}=\frac{\psi_{\mathrm{sig}}\left(W_{i},\overline{\boldsymbol{\gamma}},\overline{\boldsymbol{\alpha}},\overline{\theta}_{\mathrm{sig}}\right)}{\sqrt{n\cdot\mathrm{Var}\left(\psi_{\mathrm{sig}}\left(W_{i},\overline{\boldsymbol{\gamma}},\overline{\boldsymbol{\alpha}},\overline{\theta}_{\mathrm{sig}}\right)\right)}}.
\]
$L_{ni}$ can be viewed as a random triangular array which satisfies
\begin{eqnarray*}
\mathbb{E}\left[L_{ni}\right] & = & 0\\
\mathrm{Var}\left(L_{ni}\right) & = & \frac{1}{n}
\end{eqnarray*}
and
\[
\mathrm{Var}\left(\frac{Q_{n}-\mathbb{E}\left[Q_{n}\right]}{\sqrt{\mathrm{Var}\left(Q_{n}\right)}}\right)=1.
\]
Recall the regularity conditions
\[
\mathbb{E}\left[\left|Y_{k}-\gamma_{k}\left(X_{k}\right)\right|^{2+\eta}\mid X_{k}\right]\leq C,
\]
\[
\mathbb{E}\left|\overline{\tau}\left(Z\right)\right|^{2+\eta}<\infty,
\]
and
\[
0<\underline{e}\leq e\left(Z\right)\leq\overline{e}<1.
\]
Since $q_{n}\left(\tau\right)$ is uniformly bounded and $m_{n}\left(\tau\right)$ is dominated by $\left|\tau\right|$, there exists an integrable random variable $B\left(W\right)$ such that
\[
\left|\psi_{\mathrm{sig}}\left(W_{i}\right)\right|^{2+\eta}\leq B\left(W\right)
\]
for all $n$. Therefore,
\[
\sup_{n}\mathbb{E}\left|\psi_{\mathrm{sig}}\left(W_{i}\right)\right|^{2+\eta}<\infty.
\]
Since $\mathrm{Var}\left(\psi_{\mathrm{sig}}\left(W_{i}\right)\right)\rightarrow V$,
\begin{eqnarray*}
\sum_{i=1}^{n}\mathbb{E}\left|L_{ni}\right|^{2+\eta} & = & \sum_{i=1}^{n}\mathbb{E}\left|\frac{\psi_{\mathrm{sig}}\left(W_{i}\right)}{\sqrt{n\mathrm{Var}\left(\psi_{\mathrm{sig}}\left(W_{i}\right)\right)}}\right|^{2+\eta}\\
 & = & \frac{n\mathbb{E}\left|\psi_{\mathrm{sig}}\left(W_{i}\right)\right|^{2+\eta}}{\left(n\mathrm{Var}\left(\psi_{\mathrm{sig}}\left(W_{i}\right)\right)\right)^{1+\frac{\eta}{2}}}\\
 & = & \frac{\mathbb{E}\left|\psi_{\mathrm{sig}}\left(W_{i}\right)\right|^{2+\eta}}{n^{\frac{\eta}{2}}\left(\mathrm{Var}\left(\psi_{\mathrm{sig}}\left(W_{i}\right)\right)\right)^{1+\frac{\eta}{2}}}\\
 & \rightarrow & 0.
\end{eqnarray*}
The Lyapunov central limit theorem is applied to write
\begin{eqnarray*}
\sum_{i=1}^{n}\frac{\psi_{\mathrm{sig}}\left(w_{i}\right)}{\sqrt{n\cdot\mathrm{Var}\left(\psi_{\mathrm{sig}}\left(w_{i}\right)\right)}} & = & \frac{1}{\sqrt{n\mathrm{Var}\left(\psi_{\mathrm{sig}}\left(W_{i}\right)\right)}}\sum_{i=1}^{n}\psi_{\mathrm{sig}}\left(W_{i}\right)\\
 & \overset{d}{\rightarrow} & \mathcal{N}\left(0,1\right)
\end{eqnarray*}
which implies
\[
\frac{1}{\sqrt{n}}\sum_{i=1}^{n}\psi_{\mathrm{sig}}\left(W_{i},\overline{\boldsymbol{\gamma}},\overline{\boldsymbol{\alpha}},\overline{\theta}_{\mathrm{sig}}\right)\overset{d}{\rightarrow}\mathcal{N}\left(0,V\right)
\]
Third, let's check $\frac{1}{\sqrt{n}}\sum_{l=1}^{L}\sum_{i\in I_{l}}\left\{ \hat{R}_{1li}+\hat{R}_{2li}+\hat{R}_{3li}\right\}$.
We mostly follow the proof of Lemma 8 in \citet{Chernozhukov.et.al.2022c}.
Let $\mathcal{W}_{l}^{c}$ denote the observations not in $I_{l},$
so that $\hat{\gamma}_{l}$ and $\hat{\alpha}_{l}$ depend only on
$\mathcal{W}_{l}^{c}.$ Thus,
\begin{eqnarray*}
\mathbb{E}\left[\hat{R}_{1li}+\hat{R}_{2li}\mid\mathcal{W}_{l}^{c}\right] & = & \int\left[g\left(W_{i},\hat{\boldsymbol{\gamma}}_{l},\overline{\theta}_{\mathrm{sig}}\right)-g\left(w,\overline{\boldsymbol{\gamma}},\overline{\theta}_{\mathrm{sig}}\right)+\phi\left(W_{i},\hat{\boldsymbol{\gamma}}_{l},\overline{\boldsymbol{\alpha}}\right)-\phi\left(W_{i},\overline{\boldsymbol{\gamma}},\overline{\boldsymbol{\alpha}}\right)\right]F_{0}\left(dW_{i}\right)\\
 & = & \int\left[g\left(W_{i},\hat{\boldsymbol{\gamma}}_{l},\overline{\theta}_{\mathrm{sig}}\right)+\phi\left(W_{i},\hat{\boldsymbol{\gamma}}_{l},\overline{\boldsymbol{\alpha}}\right)\right]F_{0}\left(dW_{i}\right)\\
 & = & \mathbb{E}\left[\psi_{\mathrm{sig}}\left(w,\hat{\boldsymbol{\gamma}}_{l},\overline{\boldsymbol{\alpha}},\overline{\theta}_{\mathrm{sig}}\right)\right]\\
\mathbb{E}\left[\hat{R}_{3li}\mid\mathcal{W}_{l}^{c}\right] & = & \int\left[\phi\left(W_{i},\overline{\boldsymbol{\gamma}},\hat{\boldsymbol{\alpha}}_{l}\right)-\phi\left(W_{i},\overline{\boldsymbol{\gamma}},\overline{\boldsymbol{\alpha}}\right)\right]F_{0}\left(dW_{i}\right)\\
 & = & \int\left[\phi\left(W_{i},\overline{\boldsymbol{\gamma}},\hat{\boldsymbol{\alpha}}_{l}\right)\right]F_{0}\left(dW_{i}\right)\\
 & = & \int\left[\sum_{k=1}^{2}\phi_{k}\left(w,\overline{\gamma}_{k},\hat{\alpha}_{kl}\right)\right]F_{0}\left(dW_{i}\right)\\
 & = & \sum_{k=1}^{2}\int\hat{\alpha}_{kl}\left(X_{ki}\right)\left[Y_{ki}-\overline{\gamma}_{k}\left(X_{ki}\right)\right]F_{0}\left(dW_{i}\right)\\
 & = & 0
\end{eqnarray*}
So, we have
\[
\frac{1}{\sqrt{n}}\sum_{l=1}^{L}\sum_{i\in I_{l}}\left\{ \hat{R}_{1li}+\hat{R}_{2li}+\hat{R}_{3li}\right\} = B_{n}+A_{n,\gamma},
\]
where
\begin{eqnarray*}
    B_{n} & = & \frac{1}{\sqrt{n}}\sum_{l=1}^{L}\sum_{i\in I_{l}}\mathbb{E}\left[\hat{R}_{1li}+\hat{R}_{2li}+\hat{R}_{3li}\mid\mathcal{W}_{l}^{c}\right]\\
    & = & \frac{1}{\sqrt{n}}\sum_{l=1}^{L}\sum_{i\in I_{l}}\mathbb{E}\left[\psi_{\mathrm{sig}}\left(w,\hat{\boldsymbol{\gamma}}_{l},\overline{\boldsymbol{\alpha}},\overline{\theta}_{\mathrm{sig}}\right)\right]\\
    & = & \sum_{l=1}^{L}\frac{n_{l}}{\sqrt{n}}\mathbb{E}\left[\psi_{\mathrm{sig}}\left(w,\hat{\boldsymbol{\gamma}}_{l},\overline{\boldsymbol{\alpha}},\overline{\theta}_{\mathrm{sig}}\right)\right],
\end{eqnarray*}
and
\[
A_{n,\gamma}=\frac{1}{\sqrt{n}}\sum_{l=1}^{L}\sum_{i\in I_{l}}\left(\hat{R}_{1li}+\hat{R}_{2li}+\hat{R}_{3li}-\mathbb{E}\left[\hat{R}_{1li}+\hat{R}_{2li}+\hat{R}_{3li}\mid\mathcal{W}_{l}^{c}\right]\right).
\]
We first characterize $B_{n}$. Define
\[
\hat{d}_{l}\left(X\right)\equiv\left\{ \hat{\gamma}_{1l}\left(X\right)-\gamma_{1}\left(X\right)\right\} -\left\{ \hat{\gamma}_{2l}\left(X\right)-\gamma_{2}\left(X\right)\right\}.
\]
Since $\hat{\gamma}_{1l}\left(X\right)-\hat{\gamma}_{2l}\left(X\right)=\overline{\tau}\left(Z\right)+\hat{d}_{l}\left(X\right)$, we have
\[
m_{\mathrm{sig}}\left(W,\hat{\boldsymbol{\gamma}}_{l}\right)=m_{n}\left(\overline{\tau}\left(Z\right)+\hat{d}_{l}\left(X\right)\right).
\]
Also define
\[
a_{n}\left(t\right)\equiv m_{n}^{''}\left(t\right).
\]
Since $m_{n}\left(t\right)=t\Lambda_{n}\left(t\right)$, we have
\[
a_{n}\left(t\right)=s_{n}\Lambda_{n}\left(t\right)\left\{ 1-\Lambda_{n}\left(t\right)\right\} \left\{ 2+s_{n}t\left\{ 1-2\Lambda_{n}\left(t\right)\right\} \right\}.
\]
By the definition of the Riesz representer, we know that
\[
\mathbb{E}\left[\sum_{k=1}^{2}\overline{\alpha}_{k}\left(X_{k}\right)\left\{ \hat{\gamma}_{kl}\left(X_{k}\right)-\overline{\gamma}_{k}\left(X_{ki}\right)\right\} \right]=\mathbb{E}\left[q_{n}\left(\overline{\tau}\left(Z\right)\right)\hat{d}_{l}\left(X\right)\right].
\]
Also,
\[
\mathbb{E}\left[\sum_{k=1}^{2}\overline{\alpha}_{k}\left(X_{k}\right)\left[Y_{k}-\overline{\gamma}_{k}\left(X_{k}\right)\right]\right]=0.
\]
Therefore,
\[
\mathbb{E}\left[\psi_{\mathrm{sig}}\left(w,\hat{\boldsymbol{\gamma}}_{l},\overline{\boldsymbol{\alpha}},\overline{\theta}_{\mathrm{sig}}\right)\right]=\mathbb{E}\left[m_{n}\left\{ \overline{\tau}\left(Z\right)+\hat{d}_{l}\left(X\right)\right\} -m_{n}\left(\overline{\tau}\left(Z\right)\right)-q_{n}\left(\overline{\tau}\left(Z\right)\right)\hat{d}_{l}\left(X\right)\right].
\]
By the second-order Taylor theorem, for some $\delta_{l}\left(X\right)\in\left(0,1\right)$,
\[
m_{n}\left\{ \overline{\tau}\left(Z\right)+\hat{d}_{l}\left(X\right)\right\} -m_{n}\left(\overline{\tau}\left(Z\right)\right)-q_{n}\left(\overline{\tau}\left(Z\right)\right)\hat{d}_{l}\left(X\right)=\frac{1}{2}a_{n}\left\{ \overline{\tau}\left(Z\right)+\delta_{l}\left(X\right)\hat{d}_{l}\left(X\right)\right\} \hat{d}_{l}\left(X\right)^{2}.
\]
Hence,
\[
\mathbb{E}\left[\psi_{\mathrm{sig}}\left(w,\hat{\boldsymbol{\gamma}}_{l},\overline{\boldsymbol{\alpha}},\overline{\theta}_{\mathrm{sig}}\right)\right]=\frac{1}{2}\mathbb{E}\left[a_{n}\left\{ \overline{\tau}\left(Z\right)+\delta_{l}\left(X\right)\hat{d}_{l}\left(X\right)\right\} \hat{d}_{l}\left(X\right)^{2}\right],
\]
and the conditional mean term is expressed as
\[
B_{n}=\sum_{l=1}^{L}\frac{n_{l}}{2\sqrt{n}}\mathbb{E}\left[a_{n}\left\{ \overline{\tau}\left(Z\right)+\delta_{l}\left(X\right)\hat{d}_{l}\left(X\right)\right\} \hat{d}_{l}\left(X\right)^{2}\right].
\]
Next, we derive a worst-case upper bound. From the calculation of the second derivative,
\[
\left|a_{n}\left(t\right)\right|\leq\frac{1}{2}s_{n}
\]
uniformly over $t$. Hence,
\begin{eqnarray*}
    \left|B_{n}\right| & \leq & \sum_{l=1}^{L}\frac{n_{l}}{2\sqrt{n}}\mathbb{E}\left[\frac{1}{2}s_{n}\hat{d}_{l}\left(X\right)^{2}\right]\\
    & = & \frac{1}{4}\sqrt{n}s_{n}\sum_{l=1}^{L}\frac{n_{l}}{n}\mathbb{E}\left[\hat{d}_{l}\left(X\right)^{2}\right].
\end{eqnarray*}
Since
\[
\hat{d}_{\ell}\left(X\right)^{2}\leq2\left\{ \hat{\gamma}_{1\ell}\left(X\right)-\gamma_{1}\left(X\right)\right\} ^{2}+2\left\{ \hat{\gamma}_{2\ell}\left(X\right)-\gamma_{2}\left(X\right)\right\} ^{2},
\]
we have
\begin{eqnarray*}
    \mathbb{E}\left[\hat{d}_{\ell}\left(X\right)^{2}\right] & \leq & 2\left\Vert \hat{\gamma}_{1\ell}\left(X\right)-\gamma_{1}\left(X\right)\right\Vert ^{2}+2\left\Vert \hat{\gamma}_{2\ell}\left(X\right)-\gamma_{2}\left(X\right)\right\Vert ^{2}\\
    & = & 2\left\Vert \hat{\gamma}_{\ell}\left(X\right)-\gamma\left(X\right)\right\Vert ^{2}.
\end{eqnarray*}
Thus, we derive the worst-case bound as
\begin{eqnarray*}
    \left|B_{n}\right| & \leq & \frac{1}{2}\sqrt{n}s_{n}\sum_{l=1}^{L}\frac{n_{l}}{n}\left\Vert \hat{\gamma}_{l}-\gamma\right\Vert ^{2} \\
    & = & \frac{1}{2}\sqrt{n}s_{n}\left\Vert \hat{\gamma}-\gamma\right\Vert ^{2}.
\end{eqnarray*}
It remains to control the centered part. Note that we still have
\begin{eqnarray*}
\mathbb{E}\left[\left\{ \frac{1}{\sqrt{n}}\sum_{i\in I_{l}}\left(\hat{R}_{jli}-\mathbb{E}\left[\hat{R}_{jli}\mid\mathcal{W}_{l}^{c}\right]\right)\right\} ^{2}\mid\mathcal{W}_{l}^{c}\right] & = & \frac{n_{l}}{n}\mathrm{Var}\left(\hat{R}_{jli}\mid\mathcal{W}_{l}^{c}\right)\\
 & \leq & \mathbb{E}\left[\hat{R}_{jli}^{2}\mid\mathcal{W}_{l}^{c}\right]\\
 & = & o_{p}\left(1\right)
\end{eqnarray*}
for $j=1,2,3$ so that by the triangle and conditional Markov inequalities,
\begin{eqnarray*}
 &  & \left|\frac{1}{\sqrt{n}}\sum_{i\in I_{l}}\left(\hat{R}_{1li}+\hat{R}_{2li}+\hat{R}_{3li}-\mathbb{E}\left[\hat{R}_{1li}+\hat{R}_{2li}+\hat{R}_{3li}\mid\mathcal{W}_{l}^{c}\right]\right)\right|\\
 & \leq & \sum_{j=1}^{3}\left|\frac{1}{\sqrt{n}}\sum_{i\in I_{l}}\left(\hat{R}_{jli}-\mathbb{E}\left[\hat{R}_{jli}\mid\mathcal{W}_{l}^{c}\right]\right)\right|
\end{eqnarray*}
and for $\eta>0,$ 
\begin{eqnarray*}
 &  & \mathrm{Pr}\left(\left|\frac{1}{\sqrt{n}}\sum_{i\in I_{l}}\left(\hat{R}_{1li}+\hat{R}_{2li}+\hat{R}_{3li}-\mathbb{E}\left[\hat{R}_{1li}+\hat{R}_{2li}+\hat{R}_{3li}\mid\mathcal{W}_{l}^{c}\right]\right)\right|>3\eta\right)\\
 & \leq & \mathrm{Pr}\left(\sum_{j=1}^{3}\left|\frac{1}{\sqrt{n}}\sum_{i\in I_{l}}\left(\hat{R}_{jli}-\mathbb{E}\left[\hat{R}_{jli}\mid\mathcal{W}_{l}^{c}\right]\right)\right|>3\eta\right)\\
 & \leq & \sum_{j=1}^{3}\mathrm{Pr}\left(\left|\frac{1}{\sqrt{n}}\sum_{i\in I_{l}}\left(\hat{R}_{jli}-\mathbb{E}\left[\hat{R}_{jli}\mid\mathcal{W}_{l}^{c}\right]\right)\right|>\eta\right)\\
 & \leq & \sum_{j=1}^{3}\frac{\mathbb{E}\left[\left\{ \frac{1}{\sqrt{n}}\sum_{i\in I_{l}}\left(\hat{R}_{jli}-\mathbb{E}\left[\hat{R}_{jli}\mid\mathcal{W}_{l}^{c}\right]\right)\right\} ^{2}\mid\mathcal{W}_{l}^{c}\right]}{\eta^{2}}\\
 & \rightarrow & 0
\end{eqnarray*}
as $n\rightarrow\infty.$ Thus,
\[
\frac{1}{\sqrt{n}}\sum_{i\in I_{l}}\left(\hat{R}_{1li}+\hat{R}_{2li}+\hat{R}_{3li}-\mathbb{E}\left[\hat{R}_{1li}+\hat{R}_{2li}+\hat{R}_{3li}\mid\mathcal{W}_{l}^{c}\right]\right)=o_{p}\left(1\right)
\]
Combining the conditional mean and centered parts, we obtain
\[
\ensuremath{\frac{1}{\sqrt{n}}\sum_{l=1}^{L}\sum_{i\in I_{l}}\left\{ \hat{R}_{1li}+\hat{R}_{2li}+\hat{R}_{3li}\right\} }=B_{n}+o_{p}\left(1\right).
\]
Finally,
\[
\sqrt{n}\left(\hat{\theta}_{\mathrm{sig}}-\overline{\theta}_{\mathrm{sig}}\right)=\frac{1}{\sqrt{n}}\sum_{i=1}^{n}\psi_{\mathrm{sig}}\left(W_{i}\right)+B_{n}+o_{p}\left(1\right)
\]
where
\[
\frac{1}{\sqrt{n}}\sum_{i=1}^{n}\psi_{\mathrm{sig}}\left(W_{i}\right)\overset{d}{\rightarrow}N\left(0,V\right),
\]
with
\begin{eqnarray*}
    V & = & \mathrm{Var}\left(\overline{\tau}\left(Z_{i}\right)\mathds{1}\left\{ \overline{\tau}\left(Z_{i}\right)>0\right\} \right)+\mathbb{E}\left[\mathds{1}\left\{ \overline{\tau}\left(Z_{i}\right)>0\right\} \left\{ \frac{\sigma_{1}^{2}\left(Z_{i}\right)}{e\left(Z_{i}\right)}+\frac{\sigma_{2}^{2}\left(Z_{i}\right)}{1-e\left(Z_{i}\right)}\right\} \right]\\
    & & +\frac{1}{4}\mathbb{E}\left[\mathds{1}\left\{ \overline{\tau}\left(Z_{i}\right)=0\right\} \left\{ \frac{\sigma_{1}^{2}\left(Z_{i}\right)}{e\left(Z_{i}\right)}+\frac{\sigma_{2}^{2}\left(Z_{i}\right)}{1-e\left(Z_{i}\right)}\right\} \right].
\end{eqnarray*}
and
\[
\left|B_{n}\right|\leq\frac{1}{2}\sqrt{n}s_{n}\left\Vert \hat{\gamma}-\gamma\right\Vert ^{2}.
\]
\end{proof}

\subsection{Proposition \ref{prop:2}}
\label{app2}

\begin{proof}
Let $t=\tau\left(Z\right)$. For $t>0$,
\begin{eqnarray*}
    t\mathds{1}\left\{ \tau>0\right\} -\frac{t}{1+\exp\left(-s_{n}t\right)} & = & t-\frac{t}{1+\exp\left(-s_{n}t\right)}\\
    & = & \frac{t}{1+\exp\left(s_{n}t\right)}.
\end{eqnarray*}
For $t<0$,
\begin{eqnarray*}
    t\mathds{1}\left\{ \tau>0\right\} -\frac{t}{1+\exp\left(-s_{n}t\right)} & = & -\frac{t}{1+\exp\left(-s_{n}t\right)}\\
    & = & \frac{\left|t\right|}{1+\exp\left(s_{n}\left|t\right|\right)}.
\end{eqnarray*}
For $t=0$, the difference is zero. Therefore, for all $t\in\mathbb{R}$,
\[
t\mathds{1}\left\{ \tau>0\right\} -\frac{t}{1+\exp\left(-s_{n}t\right)}=\frac{\left|t\right|}{1+\exp\left(s_{n}\left|t\right|\right)}.
\]
Taking expectations gives
\[
\overline{\theta}-\overline{\theta}_{\mathrm{sig}}=\mathbb{E}\left[\frac{\left|\overline{\tau}\left(Z\right)\right|}{1+\exp\left(s_{n}\left|\overline{\tau}\left(Z\right)\right|\right)}\right],
\]
or equivalently,
\[
\overline{\theta}_{\mathrm{sig}}-\overline{\theta}=-\mathbb{E}\left[\frac{\left|\overline{\tau}\left(Z\right)\right|}{1+\exp\left(s_{n}\left|\overline{\tau}\left(Z\right)\right|\right)}\right].
\]
Set $u\equiv s_{n}\left|t\right|$. Then,
\[
\frac{\left|t\right|}{1+\exp\left(s_{n}\left|t\right|\right)}=\frac{1}{s_{n}}\frac{u}{1+\exp\left(u\right)}.
\]
Thus,
\[
0\leq\overline{\theta}-\overline{\theta}_{\mathrm{sig}}\leq\frac{1}{s_{n}}\sup_{u\geq0}\frac{u}{1+\exp\left(u\right)}.
\]
The supremum is obtained by solving
\[
\frac{d}{du}\left\{ \frac{u}{1+\exp\left(u\right)}\right\} =0.
\]
Since
\[
\frac{d}{du}\left\{ \frac{u}{1+\exp\left(u\right)}\right\} =\frac{1+\exp\left(u\right)-u\exp\left(u\right)}{\left\{ 1+\exp\left(u\right)\right\} ^{2}},
\]
the first-order condition is
\[
1+\exp\left(u\right)-u\exp\left(u\right)=0
\]
or
\[
\left(u-1\right)\exp\left(u\right)=1.
\]
Let $v=u-1$. Then,
\[
v\exp\left(v\right)=e^{-1},
\]
so
\[
v=W\left(e^{-1}\right)
\]
and
\[
u=1+W\left(e^{-1}\right).
\]
At this maximizer,
\[
\exp\left(u\right)=\frac{1}{u-1}.
\]
Therefore,
\begin{eqnarray*}
    \sup_{u\geq0}\frac{u}{1+\exp\left(u\right)} & = & \frac{u}{1+\frac{1}{u-1}}\\
    & = & u-1\\
    & = & W\left(e^{-1}\right).
\end{eqnarray*}
Hence,
\[
0\leq\overline{\theta}-\overline{\theta}_{\mathrm{sig}}\leq\frac{W\left(e^{-1}\right)}{s_{n}}.
\]
\end{proof}

\subsection{Theorem \ref{thm:oracle-bias-aware}}
\label{app5}

\begin{proof}[Proof of Theorem \ref{thm:oracle-bias-aware}]
First, observe that
\[
\begin{aligned}
&\widehat\theta_{\mathrm{sig}}-\bar\theta
-\left(
\frac{B_n}{\sqrt n}
+\bar\theta_{\mathrm{sig}}-\bar\theta
\right) \\
&\qquad =
\widehat\theta_{\mathrm{sig}}
-\bar\theta_{\mathrm{sig}}
-\frac{B_n}{\sqrt n}.
\end{aligned}
\]
Therefore, applying Proposition 1 at \(s_n=s_n^*\) gives
\[
\sqrt n
\left[
\widehat\theta_{\mathrm{sig}}-\bar\theta
-\left(
\frac{B_n}{\sqrt n}
+\bar\theta_{\mathrm{sig}}-\bar\theta
\right)
\right]
=
\frac{1}{\sqrt n}
\sum_{i=1}^n\psi_{\mathrm{sig}}(W_i)
+o_p(1).
\]
Because
\[
\frac{1}{\sqrt n}
\sum_{i=1}^n\psi_{\mathrm{sig}}(W_i)
\overset{d}{\longrightarrow}N(0,V)
\]
and
\[
\sqrt n\,
\operatorname{se}(\widehat\theta_{\mathrm{sig}})
\overset{p}{\longrightarrow}\sqrt V,
\]
Slutsky's theorem yields
\[
\frac{
\widehat\theta_{\mathrm{sig}}-\bar\theta
-\left(
B_n/\sqrt n+\bar\theta_{\mathrm{sig}}-\bar\theta
\right)
}{
\operatorname{se}(\widehat\theta_{\mathrm{sig}})
}
\overset{d}{\longrightarrow}N(0,1).
\]
It remains to establish the bounds on the combined drift. By Proposition \ref{prop:1} and
\[
s_n^*
=
\frac{\sqrt{2W(e^{-1})}}
{\|\widehat\gamma-\overline{\gamma}\|},
\]
we have
\[
\begin{aligned}
\frac{|B_n|}{\sqrt n}
&\leq
\frac{1}{2}s_n^*
\|\widehat\gamma-\overline{\gamma}\|^2 \\
&=
\sqrt{\frac{W(e^{-1})}{2}}\,
\|\widehat\gamma-\overline{\gamma}\|.
\end{aligned}
\]
Proposition \ref{prop:2} similarly gives
\[
-\frac{W(e^{-1})}{s_n^*}
\leq
\bar\theta_{\mathrm{sig}}-\bar\theta
\leq 0,
\]
where
\[
\frac{W(e^{-1})}{s_n^*}
=
\sqrt{\frac{W(e^{-1})}{2}}\,
\|\widehat\gamma-\overline{\gamma}\|.
\]
Combining these bounds gives
\[
-\sqrt{2W(e^{-1})}\,
\|\widehat\gamma-\overline{\gamma}\|
\leq
\frac{B_n}{\sqrt n}
+\bar\theta_{\mathrm{sig}}-\bar\theta
\leq
\sqrt{\frac{W(e^{-1})}{2}}\,
\|\widehat\gamma-\overline{\gamma}\|.
\]
Consequently,
\[
\left|
\frac{B_n}{\sqrt n}
+\bar\theta_{\mathrm{sig}}-\bar\theta
\right|
\leq
\sqrt{2W(e^{-1})}\,
\|\widehat\gamma-\overline{\gamma}\|.
\]
This completes the proof.
\end{proof}

\bibliographystyle{apalike} 
\bibliography{DML_GP}

@Article{Andrews2000,
  author    = {Donald W. K. Andrews},
  journal   = {Econometrica},
  title     = {Inconsistency of the Bootstrap When a Parameter is on the Boundary of the Parameter Space},
  year      = {2000},
  issn      = {00129682, 14680262},
  number    = {2},
  pages     = {399--405},
  volume    = {68},
  publisher = {[Wiley, Econometric Society]},  
}

@Article{Andrews.et.al2023,
  author   = {Andrews, Isaiah and Kitagawa, Toru and McCloskey, Adam},
  journal  = {The Quarterly Journal of Economics},
  title    = {Inference on Winners},
  year     = {2024},
  issn     = {0033-5533},
  month    = {09},
  number   = {1},
  pages    = {305-358},
  volume   = {139},  
}

@Article{Armstrong.and.Kolesar2020,
  author   = {Armstrong, Timothy B. and Kolesár, Michal},
  journal  = {Quantitative Economics},
  title    = {Simple and honest confidence intervals in nonparametric regression},
  year     = {2020},
  number   = {1},
  pages    = {1-39},
  volume   = {11},  
}

@article{Athey.and.Wager2019,
  title={Estimating treatment effects with causal forests: An application},
  author={Athey, Susan and Wager, Stefan},
  journal={Observational studies},
  volume={5},
  number={2},
  pages={37--51},
  year={2019},
  publisher={University of Pennsylvania Press}
}

@article{Athey.and.Wager2021,
author = {Athey, Susan and Wager, Stefan},
title = {Policy Learning With Observational Data},
journal = {Econometrica},
volume = {89},
number = {1},
pages = {133-161},
year = {2021}
}

@Article{Balke.and.Pearl.1997,
    author = {Alexander Balke and Judea Pearl},
    title = {Bounds on Treatment Effects from Studies with Imperfect Compliance},
    journal = {Journal of the American Statistical Association},
    volume = {92},
    number = {439},
    pages = {1171--1176},
    year = {1997},
    publisher = {ASA Website},
}

@Article{Belloni.and.Chernozhukov2013,
  author    = {Alexandre Belloni and Victor Chernozhukov},
  journal   = {Bernoulli},
  title     = {{Least squares after model selection in high-dimensional sparse models}},
  year      = {2013},
  number    = {2},
  pages     = {521 -- 547},
  volume    = {19},
  publisher = {Bernoulli Society for Mathematical Statistics and Probability},  
}

@Article{Bickel.et.al.2009,
  author    = {Peter J. Bickel and Ya’acov Ritov and Alexandre B. Tsybakov},
  journal   = {The Annals of Statistics},
  title     = {{Simultaneous analysis of Lasso and Dantzig selector}},
  year      = {2009},
  number    = {4},
  pages     = {1705 -- 1732},
  volume    = {37},
  publisher = {Institute of Mathematical Statistics},  
}

@Article{Bloom.et.al.1997,
  author    = {Howard S. Bloom and Larry L. Orr and Stephen H. Bell and George Cave and Fred Doolittle and Winston Lin and Johannes M. Bos},
  journal   = {The Journal of Human Resources},
  title     = {The Benefits and Costs of JTPA Title II-A Programs: Key Findings from the National Job Training Partnership Act Study},
  year      = {1997},
  issn      = {0022166X},
  number    = {3},
  pages     = {549--576},
  volume    = {32},
  publisher = {[University of Wisconsin Press, Board of Regents of the University of Wisconsin System]},  
}

@Article{Chen.et.al.2003,
  author    = {Chen, Xiaohong and Linton, Oliver and Van Keilegom, Ingrid},
  journal   = {Econometrica},
  title     = {Estimation of Semiparametric Models When the Criterion Function Is Not Smooth},
  year      = {2003},
  issn      = {00129682, 14680262},
  number    = {5},
  pages     = {1591--1608},
  volume    = {71},
  publisher = {[Wiley, Econometric Society]},  
}

@misc{Chen.et.al.2025,
      title={Inference on Welfare and Value Functionals under Optimal Treatment Assignment}, 
      author={Xiaohong Chen and Zhenxiao Chen and Wayne Yuan Gao},
      year={2025},
      eprint={2510.25607},
      archivePrefix={arXiv},
      primaryClass={econ.EM},      
}

@Article{Chen.and.Pouzo.2012,
  author    = {Xiaohong Chen and Demian Pouzo},
  journal   = {Econometrica},
  title     = {ESTIMATION OF NONPARAMETRIC CONDITIONAL MOMENT MODELS WITH POSSIBLY NONSMOOTH GENERALIZED RESIDUALS},
  year      = {2012},
  issn      = {00129682, 14680262},
  number    = {1},
  pages     = {277--321},
  volume    = {80},
  publisher = {[Wiley, Econometric Society]},  
}

@Article{Chernozhukov.et.al.2017,
  author  = {Chernozhukov, Victor and Chetverikov, Denis and Demirer, Mert and Duflo, Esther and Hansen, Christian and Newey, Whitney K.},
  journal = {American Economic Review},
  title   = {Double/Debiased/Neyman Machine Learning of Treatment Effects},
  year    = {2017},
  month   = {May},
  number  = {5},
  pages   = {261-65},
  volume  = {107},  
}

@Article{Chernozhukov.et.al.2018,
  author   = {Chernozhukov, Victor and Chetverikov, Denis and Demirer, Mert and Duflo, Esther and Hansen, Christian and Newey, Whitney K. and Robins, James},
  journal  = {The Econometrics Journal},
  title    = {{Double/debiased machine learning for treatment and structural parameters}},
  year     = {2018},
  issn     = {1368-4221},
  month    = {01},
  number   = {1},
  pages    = {C1-C68},
  volume   = {21},  
}

@Article{Chernozhukov.et.al.2022a,
  author   = {Chernozhukov, Victor and Newey, Whitney K. and Singh, Rahul},
  journal  = {Econometrica},
  title    = {Automatic Debiased Machine Learning of Causal and Structural Effects},
  year     = {2022},
  number   = {3},
  pages    = {967-1027},
  volume   = {90},  
}

@Article{Chernozhukov.et.al.2022b,
  author   = {Chernozhukov, Victor and Newey, Whitney K. and Singh, Rahul},
  journal  = {The Econometrics Journal},
  title    = {{Debiased machine learning of global and local parameters using regularized Riesz representers}},
  year     = {2022},
  issn     = {1368-4221},
  month    = {04},
  number   = {3},
  pages    = {576-601},
  volume   = {25},  
}

@Article{Chernozhukov.et.al.2022c,
  author   = {Chernozhukov, Victor and Escanciano, Juan Carlos and Ichimura, Hidehiko and Newey, Whitney K. and Robins, James M.},
  journal  = {Econometrica},
  title    = {Locally Robust Semiparametric Estimation},
  year     = {2022},
  number   = {4},
  pages    = {1501-1535},
  volume   = {90},  
}

@misc{Chernozhukov.et.al.2024a,
      title={Applied Causal Inference Powered by ML and AI}, 
      author={Victor Chernozhukov and Christian Hansen and Nathan Kallus and Martin Spindler and Vasilis Syrgkanis},
      year={2024},
      eprint={2403.02467},
      archivePrefix={arXiv},
      primaryClass={econ.EM},      
}

@Article{Chernozhukov.et.al.2024b,
      title={Adversarial Estimation of Riesz Representers},
      journal  = {Journal of the American Statistical Association},
      author={Victor Chernozhukov and Whitney K. Newey and Rahul Singh and Vasilis Syrgkanis},
      year={2026},
      pages = {1398-1409},
      volume = {121},
}

@Article{Christensen.et.al.2023,
  author        = {Timothy Christensen and Hyungsik Roger Moon and Frank Schorfheide},
  title         = {Optimal Decision Rules when Payoffs are Partially Identified},
  journal  = {The Review of Economic Studies},
  year     = {2026},
  volume   = {rdag017},
}

@Article{Fang.and.Santos.2018,
  author   = {Fang, Zheng and Santos, Andres},
  journal  = {The Review of Economic Studies},
  title    = {{Inference on Directionally Differentiable Functions}},
  year     = {2019},
  issn     = {0034-6527},
  month    = {09},
  number   = {1},
  pages    = {377-412},
  volume   = {86},  
}

@Article{Hong.and.Li.2018,
  author   = {Han Hong and Jessie Li},
  journal  = {Journal of Econometrics},
  title    = {The numerical delta method},
  year     = {2018},
  issn     = {0304-4076},  
  number   = {2},
  pages    = {379-394},
  volume   = {206},  
}

@Article{Hirano.and.Porter.2012,
  author    = {Keisuke Hirano and Jack R. Porter},
  journal   = {Econometrica},
  title     = {IMPOSSIBILITY RESULTS FOR NONDIFFERENTIABLE FUNCTIONALS},
  year      = {2012},
  issn      = {00129682, 14680262},
  number    = {4},
  pages     = {1769--1790},
  volume    = {80},
  publisher = {[Wiley, Econometric Society]},  
}

@Article{Hirshberg.and.Wager.2021,
  author    = {David A. Hirshberg and Stefan Wager},
  journal   = {The Annals of Statistics},
  title     = {{Augmented minimax linear estimation}},
  year      = {2021},
  number    = {6},
  pages     = {3206 -- 3227},
  volume    = {49},
  publisher = {Institute of Mathematical Statistics},  
}

@Article{Horowitz1992,
  author    = {Joel L. Horowitz},
  journal   = {Econometrica},
  title     = {A Smoothed Maximum Score Estimator for the Binary Response Model},
  year      = {1992},
  issn      = {00129682, 14680262},
  number    = {3},
  pages     = {505--531},
  volume    = {60},
  publisher = {[Wiley, Econometric Society]},  
}

@Article{Imai.et.al.2010,
  author  = {Imai, Kosuke and Keele, Luke and Tingley, Dustin},
  journal = {Psychological Methods},
  title   = {A General Approach to Causal Mediation Analysis},
  year    = {2010},
  month   = dec,
  number  = {4},
  pages   = {309--334},
  volume  = {15},
}

@Article{Kitagawa.and.Tetenov2018,
  author   = {Kitagawa, Toru and Tetenov, Aleksey},
  journal  = {Econometrica},
  title    = {Who Should Be Treated? Empirical Welfare Maximization Methods for Treatment Choice},
  year     = {2018},
  number   = {2},
  pages    = {591-616},
  volume   = {86},  
}

@Article{Kitagawa.et.al2020,
  author   = {Toru Kitagawa and José Luis {Montiel Olea} and Jonathan Payne and Amilcar Velez},
  journal  = {Journal of Econometrics},
  title    = {Posterior distribution of nondifferentiable functions},
  year     = {2020},
  issn     = {0304-4076},
  number   = {1},
  pages    = {161-175},
  volume   = {217},  
}

@misc{Klosin2021,
      title={Automatic Double Machine Learning for Continuous Treatment Effects}, 
      author={Sylvia Klosin},
      year={2021},
      eprint={2104.10334},
      archivePrefix={arXiv},
      primaryClass={econ.EM},      
}

@Article{Künzel.el.al.2019,
    author = {Sören R. Künzel  and Jasjeet S. Sekhon  and Peter J. Bickel  and Bin Yu },
    title = {Metalearners for estimating heterogeneous treatment effects using machine learning},
    journal = {Proceedings of the National Academy of Sciences},
    volume = {116},
    number = {10},
    pages = {4156-4165},
    year = {2019},    
}

@Article{Levis.et.al.2023,
      title={Covariate-assisted bounds on causal effects with instrumental variables}, 
      author={Alexander W. Levis and Matteo Bonvini and Zhenghao Zeng and Luke Keele and Edward H. Kennedy},
      year={2025},
      journal = {Journal of the Royal Statistical Society Series B: Statistical Methodology},      
      volume = {87},
    number = {5},
    pages = {1508-1527},
}

@Article{Luedtke.and.van.der.Laan2016,
  author    = {Alexander R. Luedtke and Mark J. van der Laan},
  journal   = {The Annals of Statistics},
  title     = {{Statistical inference for the mean outcome under a possibly non-unique optimal treatment strategy}},
  year      = {2016},
  number    = {2},
  pages     = {713 -- 742},
  volume    = {44},
  publisher = {Institute of Mathematical Statistics},  
}

@Article{Manski2004,
author = {Manski, Charles F.},
title = {Statistical Treatment Rules for Heterogeneous Populations},
journal = {Econometrica},
volume = {72},
number = {4},
pages = {1221-1246},
year = {2004}
}

@Article{Newey1994,
  author    = {Whitney K. Newey},
  journal   = {Econometrica},
  title     = {The Asymptotic Variance of Semiparametric Estimators},
  year      = {1994},
  issn      = {00129682, 14680262},
  number    = {6},
  pages     = {1349--1382},
  volume    = {62},
  publisher = {[Wiley, Econometric Society]},  
}

@article{Robins.et.al1994,
 ISSN = {01621459, 1537274X},
 author = {James M. Robins and Andrea Rotnitzky and Lue Ping Zhao},
 journal = {Journal of the American Statistical Association},
 number = {427},
 pages = {846--866},
 publisher = {[American Statistical Association, Taylor & Francis, Ltd.]},
 title = {Estimation of Regression Coefficients When Some Regressors Are Not Always Observed},
 urldate = {2025-02-20},
 volume = {89},
 year = {1994}
}

@Article{Rudelson.and.Zhou2013,
  author   = {Rudelson, Mark and Zhou, Shuheng},
  journal  = {IEEE Transactions on Information Theory},
  title    = {Reconstruction From Anisotropic Random Measurements},
  year     = {2013},
  number   = {6},
  pages    = {3434-3447},
  volume   = {59},
}

@Article{Semenova2023a,
  author   = {Vira Semenova},
  journal  = {Journal of Econometrics},
  title    = {Generalized Lee bounds},
  year     = {2025},  
  pages    = {106055},
  volume   = {251},  
}

@misc{Semenova2024,
      title={Debiased Machine Learning of Aggregated Intersection Bounds and Other Causal Parameters}, 
      author={Vira Semenova},
      year={2025},
      eprint={2303.00982},
      archivePrefix={arXiv},
      primaryClass={econ.EM},      
}

@Article{Semenova.and.Chernozhukov2020,
  author   = {Semenova, Vira and Chernozhukov, Victor},
  journal  = {The Econometrics Journal},
  title    = {{Debiased machine learning of conditional average treatment effects and other causal functions}},
  year     = {2021},
  issn     = {1368-4221},
  month    = {08},
  number   = {2},
  pages    = {264-289},
  volume   = {24},  
}

@misc{Whitehouse.et.al.2026,
      title={Inference on Optimal Policy Values and Other Irregular Functionals via Softmax Smoothing}, 
      author={Justin Whitehouse and Qizhao Chen and Morgane Austern and Vasilis Syrgkanis},
      year={2026},
      eprint={2507.11780},
      archivePrefix={arXiv},
      primaryClass={econ.EM},      
}

@article{Zhou.et.al2017,
author = {Xin Zhou and Nicole Mayer-Hamblett and Umer Khan and Michael R. Kosorok},
title = {Residual Weighted Learning for Estimating Individualized Treatment Rules},
journal = {Journal of the American Statistical Association},
volume = {112},
number = {517},
pages = {169--187},
year = {2017},
publisher = {ASA Website},
}

\end{document}